\documentclass[twocolumn,aps,prd,reprint,nofootinbib,url = false,longbibliography]{revtex4-1}

\usepackage[utf8]{inputenc}
\usepackage[english]{babel}
\usepackage{dsfont}
\usepackage{amsmath}
\usepackage{ mathrsfs }
\usepackage{amssymb}
\usepackage{graphicx}%
\usepackage{chemarr}
\usepackage{dcolumn}
\usepackage{bm}
\usepackage{amsmath}
\usepackage{varioref}
\usepackage{comment}
\usepackage[bottom]{footmisc}

\usepackage{physics}
\usepackage{listings}

\newcommand{\RN}[1]{%
  \textup{\uppercase\expandafter{\romannumeral#1}}%
}
\newcommand*\dif{\mathop{}\!\mathrm{d}}

\usepackage{graphicx}

\newcolumntype{C}{>{$}c<{$}}
\AtBeginDocument{
\heavyrulewidth=.08em
\lightrulewidth=.05em
\cmidrulewidth=.03em
\belowrulesep=.65ex
\belowbottomsep=0pt
\aboverulesep=.4ex
\abovetopsep=0pt
\cmidrulesep=\doublerulesep
\cmidrulekern=.5em
\defaultaddspace=.5em
}

\usepackage{floatrow}
\usepackage{graphicx}

\usepackage{booktabs}

\usepackage[labelformat=simple,subrefformat=parens,labelformat=parens]{subfig}
\usepackage{caption}
\newcommand{\com}[1]{\textcolor{blue}{#1}}

\begin{document}

\title{Non-equilibrium wet-dry cycling acts as a catalyst for chemical reactions}

\author{Ivar Svalheim Haugerud, Pranay Jaiswal, Christoph A. Weber}
\email[]{christoph.weber@physik.uni-augsburg.de}

\affiliation{Faculty of Mathematics, Natural Sciences, and Materials Engineering: Institute of Physics, University of Augsburg, Universit\"atsstra{\ss}e\ 1, 86159 Augsburg, Germany}


\begin{abstract}
\textbf{Abstract:}
Recent experimental studies suggest that wet-dry cycles and coexisting phases can each strongly alter chemical processes. The mechanisms of why and to which degree chemical processes are altered when subject to evaporation and condensation are unclear. To close this gap, we developed a theoretical framework for non-dilute chemical reactions subject to non-equilibrium conditions of evaporation {and} condensation. We find that such conditions can change the half-time of the product's yield by more than an order of magnitude, depending on the substrate-solvent interaction. We show that the cycle frequency strongly affects the chemical turnover when maintaining the system out of equilibrium by wet-dry cycles. There {exists a} resonance behavior in the cycle frequency where the turnover is maximal. This resonance behavior enables wet-dry cycles to select specific chemical reactions suggesting a potential mechanism for chemical evolution in prebiotic soups at early Earth.
\end{abstract}
\maketitle

The presence of catalysts severely alters the speed of chemical reactions.
In contrast to reactants, they are not converted during the turnover of a substrate to a product. 
Catalysts affect the activation energy $\Delta E$ without changing the thermodynamic free energies of the substrate and product state \cite{masel_chemical_nodate}.
The use of catalysts ensures economic efficiency of most applications in chemical and biomolecular engineering~\cite{singh_catalysis_2014,werner_ionic_2010,blaser_role_1999}.
In living systems, catalytic activity is mediated by enzymes. These enzymes are specialized proteins involved in almost all metabolic processes in the cell~\cite{alberty:2003,srere_complexes_1987,underkofler_production_1958}. Enzymes do not solely maximize the speed of chemical reactions; they are optimal concerning reaction speed and specificity~\cite{eigen_elementary_1963,eigen_new_1968}, i.e., they catalyze only specific reactions in a mixture of extremely many reacting components. However, this optimal catalytic property for life had to develop during evolution~\cite{kauffman_approaches_2011,oparin_origin_1965,orgel_origin_1994}. In other words, there were likely no enzymes at the molecular origin of life. It remains one of the mysteries of the molecular origin of life how functional products such as self-replicating RNA strands could have formed despite activation barriers significantly larger than  $k_BT$~\cite{hay_barrier_2010,claeyssens_high-accuracy_2006}.

\begin{figure}[b]
    \centering
     \makebox[\textwidth][c]{
     \sidesubfloat[]{\includegraphics[width=0.9\textwidth]{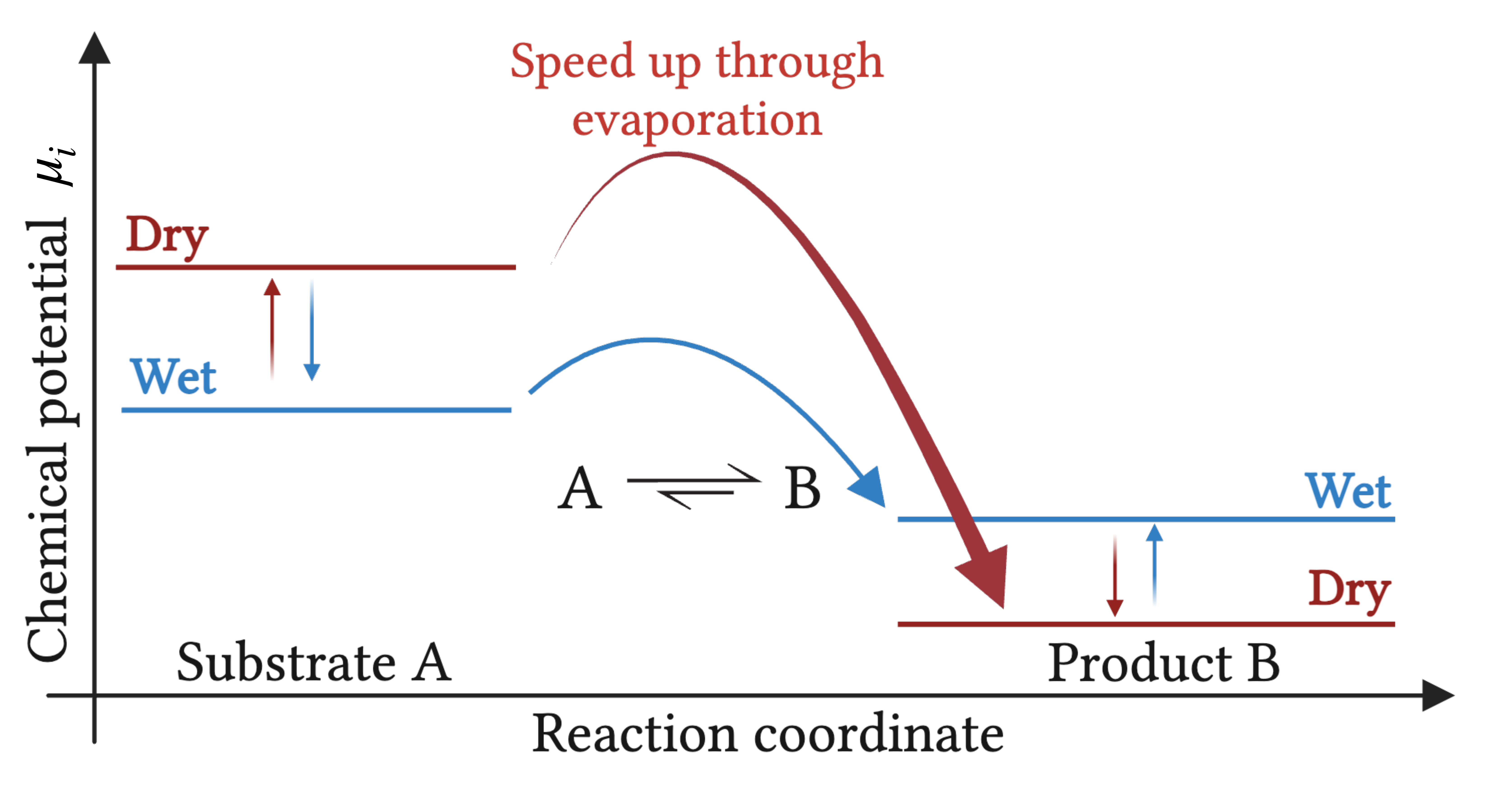}\label{fig:1a}}}
     \makebox[\textwidth][c]{
     \sidesubfloat[]{\includegraphics[width=0.40\textwidth]{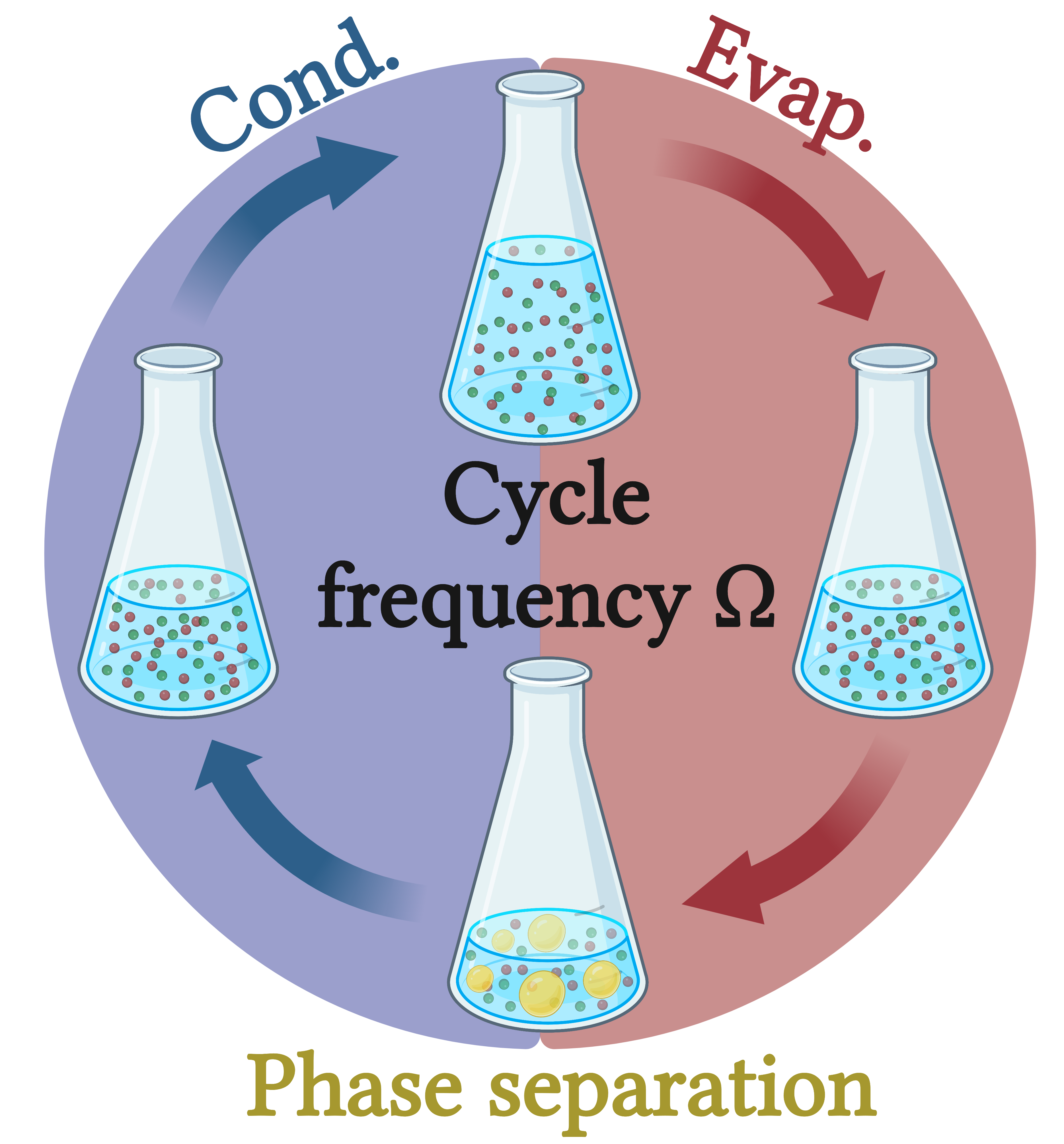}\label{fig:1b}}
    \sidesubfloat[]{\includegraphics[width=0.45\textwidth]{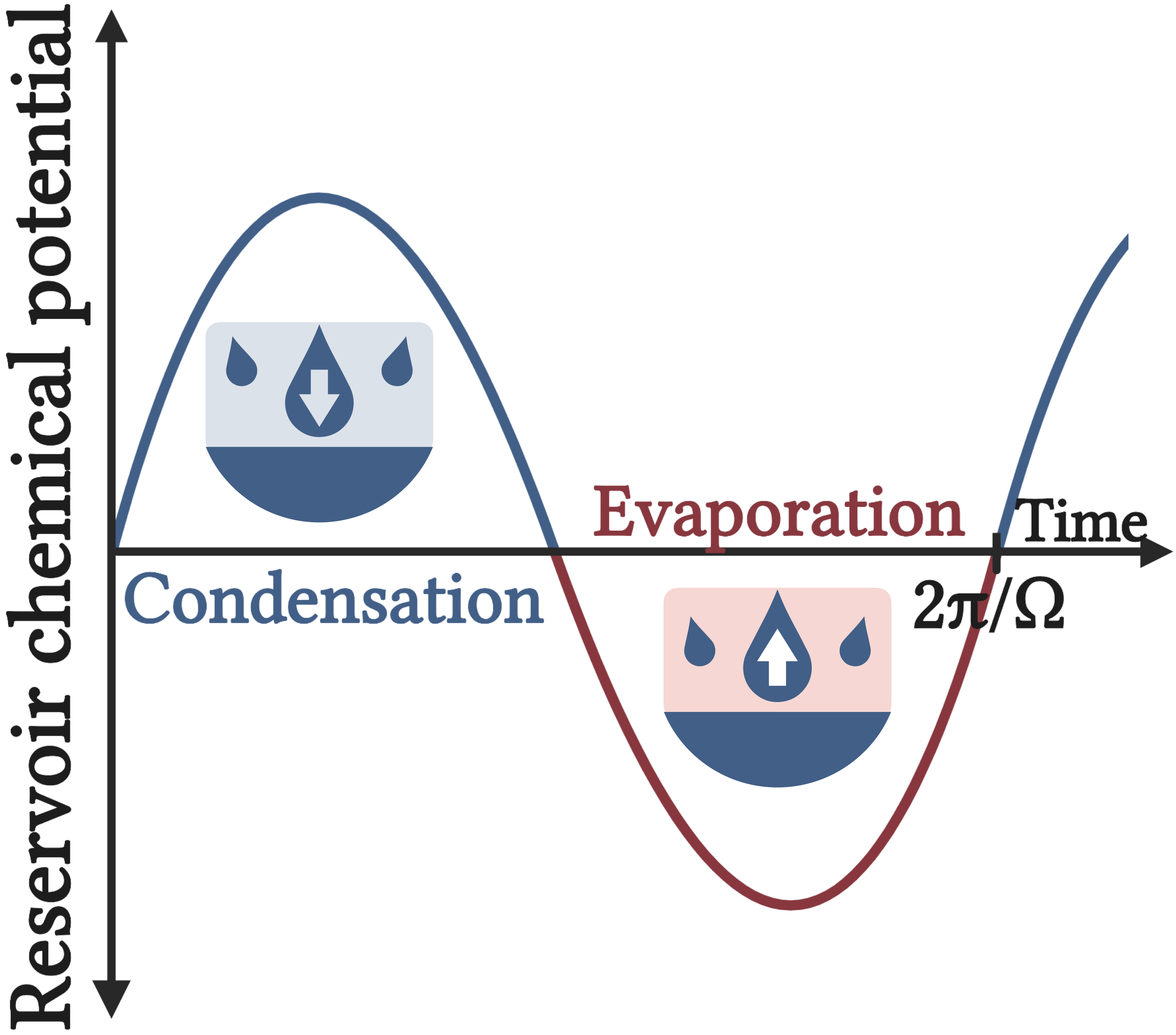}\label{fig:1c}}
    }
\caption{\textbf{Wet-dry cycles in chemically reacting systems:}
(a) Wet-dry cycles can enhance chemical reaction rates by increasing the chemical potential of the substrate $\mu_A$ and decreasing the chemical potential $\mu_B$ of the product.
(b) Wet-dry cycles lead to non-dilute conditions, which affect chemical reactions and may induce liquid-liquid phase separation. (c) The cyclic behavior is included through an oscillating reservoir  with periods of evaporation and condensation.
   } 
   \label{fig:1}
\end{figure}

An alternative perspective on how to speed up chemical reactions is through non-equilibrium conditions~\cite{mast_escalation_2013,morasch_heated_2019,baaske_extreme_2007,bartolucci_selection_2022,ianeselli_non-equilibrium_2022,matreux_heat_2021,busiello_dissipation-driven_2021}. In fact, non-equilibrium applies to many processes in chemical engineering, almost all processes in living cells and prebiotic soup at early Earth. The reason is that these are all open systems that can exchange matter and entropy with their environment, termed dissipate systems~\cite{nicolis_dissipative_1986, england_dissipative_2015,brogliato_dissipative_2020}. 
Continuous dissipation is often achieved by cycles of the system's control parameters, such as temperature or chemical potentials. Recently studied cases are wet-dry cycles~\cite{maguire_physicochemical_2021,fares_impact_2020,becker_wet-dry_2018,tekin_prebiotic_2022,dass_rna_2023,higgs_effect_2016}.

The ability of wet-dry cycles to affect the synthesis of compounds that are also relevant for the molecular origin of life is progressively backed by experiments~\citep{forsythe_ester-mediated_2015,mamajanov_ester_2014,becker_wet-dry_2018}. 
Wet-dry cycles have been applied to catalyze the formation  of polyacids~\citep{weber_thermal_1989}, esters~\citep{forsythe_ester-mediated_2015,mamajanov_ester_2014}, oligopeptides~\citep{lahav_peptide_1978,Rodriguez-garcia_formation_2015,fox_thermal_1958}, nucleosides~\citep{becker_wet-dry_2018}, polynucleotides~\citep{rajamani_lipid-assisted_2008,da_silva_salt-promoted_2015}, polyethersulfone~\cite{fang_effect_1994}, phosphorylation~\citep{maguire_physicochemical_2021} and lipid structures~\citep{rajamani_lipid-assisted_2008}.
Up to now, the mechanisms of how wet-dry cycles affect chemical processes, particularly in comparison to catalysts, remain elusive.

Unraveling such mechanisms is complex since reducing the solvent amount through evaporation generally leads to non-dilute conditions. As a result,  non-linear thermodynamic contributions to chemical activities emerge in addition to altered kinetic activation barriers.  
Additionally, non-dilute systems can induce phase transitions, such as liquid-liquid phase separation~\cite{weber_physics_2019,kato_phase_2021,hyman_liquid-liquid_2014}, gel formation~\cite{blankschteinTheoryPhaseSeparation1985,semenovThermoreversibleGelationSolutions1998,  deviriEquilibriumSizeDistribution2020,bartolucci2023interplay}, and solid precipitation~\cite{lewis2010review, agarwal2014solute}. Recent experiments show that coacervate formation can be triggered by wet-dry cycles~\cite{fares_impact_2020}. 
To interpret the existing experimental studies on the effects of wet-dry cycles on the synthesis of chemical buildings
and to decipher the underlying physio-chemical mechanisms, a theoretical framework is lacking.

This work proposes a kinetic theory based on thermodynamic principles for a non-dilute chemically reacting mixture coupled to a cyclic wet-dry reservoir that drives the system out of equilibrium; see Fig.~\ref{fig:1}b,c. We show to what extent and under which conditions both evaporation and condensation of solvent can speed up or slow down rates of chemical reactions.
In contrast to catalysts that solely affect activation barriers, we find that solvent evaporation and condensation can significantly alter the chemical potentials of substrate and product once the system is driven toward non-dilute conditions, illustrated in Fig.~\ref{fig:1a}. 
We also study how wet-dry cycles allow for continuous chemical activity by keeping the system out of equilibrium. For this case, a resonance cycling frequency exists for each evaporation/condensation flux at which the chemical turnover is maximal. 
This resonance behavior provides a selection pressure for specific chemical reaction networks suggesting wet-dry cycles as potential catalyst supplements to kick-start the molecular evolution of life~\cite{carrea_polyamino_2005,eijsink_directed_2005,schulz_evolution_2022}.

\section*{Theory for wet-dry cycles with chemical reactions}

We consider a mixture of volume V composed of $M+1$ components, each of particle number $N_i$ ($i=0,1,...,M$) in the Gibbs ($T$-$P$-$N_i$) ensemble with the Gibbs free energy $G(T,P,N_i)$. 
Evaporation and condensation of molecule $i$ is governed by the difference between the bulk chemical potential $\mu_i=  \left({\partial G}/{\partial N_i}\right)_{T, p, N_{j\neq i}}$, and the reservoir  chemical potential $\mu_i^\text{r}$.
In all our studies, the chemical potential of the reservoir $\mu_i^\text{r}$ is a control parameter that can be constant or vary in time.
The evaporation and condensation flux $h_i$ describes the number of molecules passing through the surface area $A$ per unit time \cite{weber_physics_2019}:
\begin{align}
  \frac{h_i}{A}  &= k_{\text{e},i} \left[\exp{\frac{\mu_i^\text{r}}{k_\text{B}T}}-\exp{\frac{\mu_i}{k_\text{B}T}}\right] \, , \label{eq:evap}
\end{align}
where $k_\text{B}T$ denotes the thermal energy. Here, $k_{\text{e},i}$ is the kinetic evaporation/condensation coefficient for component $i$. Equilibrium is reached when the chemical potentials between the reservoir and the bulk are equal, $\mu_i=\mu_i^\text{r}$, implying no exchange between the bulk and the reservoir,
$h_i=0$. 
Note that Eq.~\eqref{eq:evap} considers a symmetric form, $h_i=h_i^\text{cond}-h_i^\text{evap}$, of the condensation flux, $h_i^\text{cond}/A=k_{\text{e},i} \exp{{\mu_i^\text{r}}/{k_\text{B}T}}$, and evaporation flux, $h_i^\text{evap}/A =k_{\text{e},i} \exp{{\mu_i^\text{}}/{k_\text{B}T}}$, without the loss of generality. The key property is that detailed balance of the rates, $h_i^\text{cond}/h_i^\text{evap}=\exp{(\mu_i^\text{r}-\mu_i^\text{})/{k_\text{B}T}}$, is satisfied.

\begin{figure*}[tb]
    \centering
     \makebox[\textwidth][c]{
     \sidesubfloat[]{\includegraphics[height=0.3\textwidth]{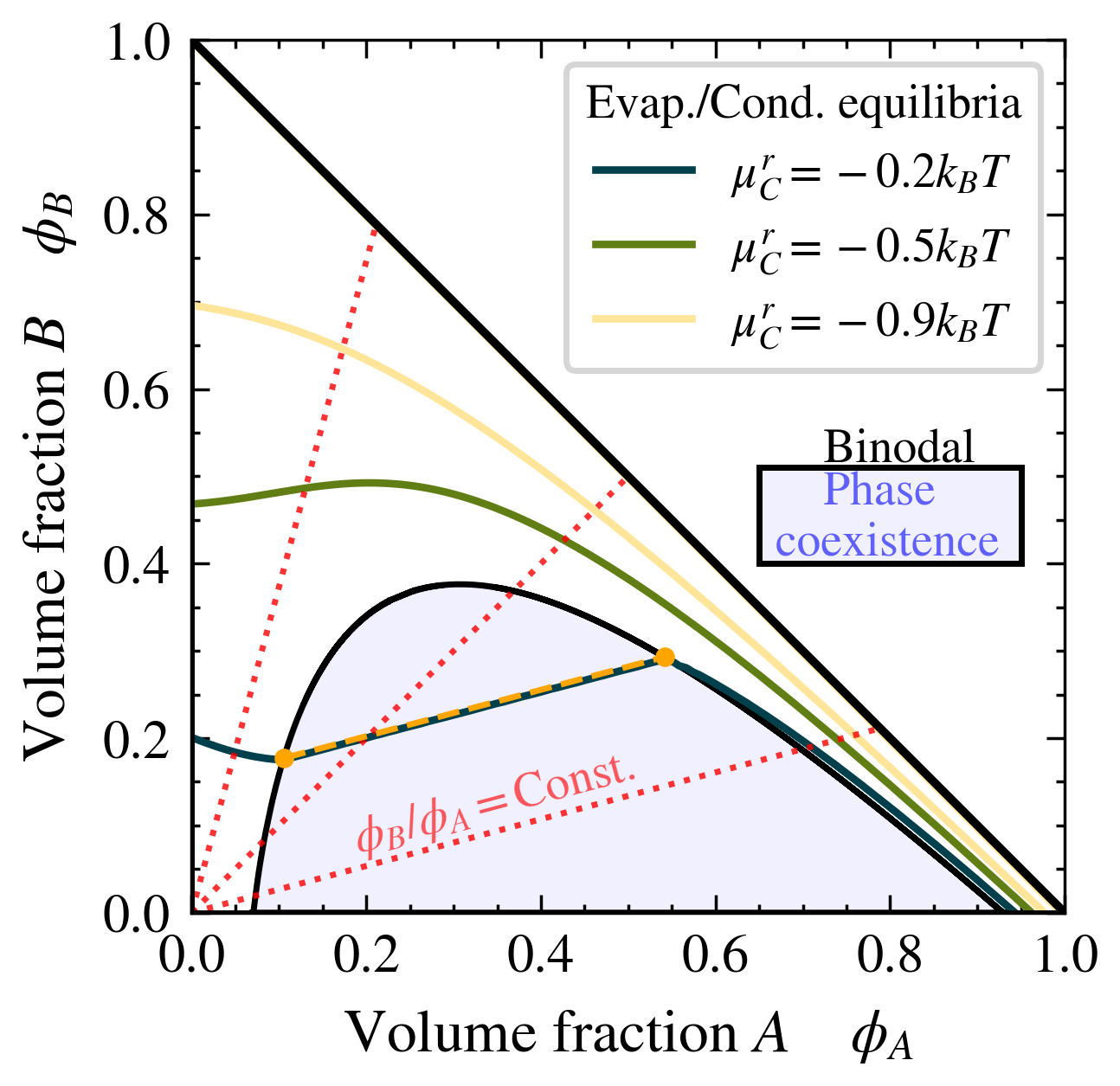}\label{fig:2a}}
    \sidesubfloat[]{\includegraphics[height=0.3\textwidth]{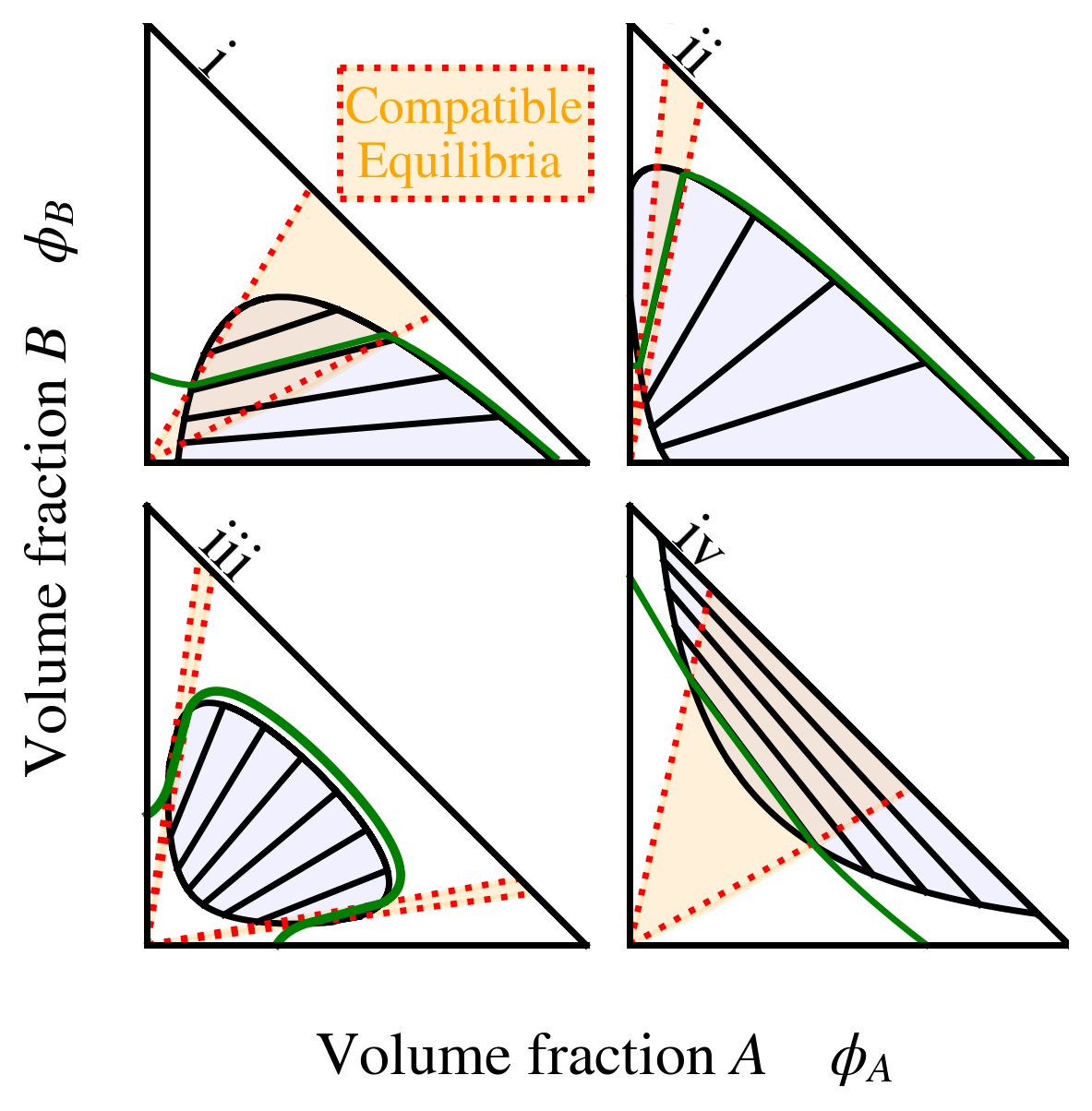}\label{fig:2b}}
     \sidesubfloat[]{\includegraphics[height=0.3\textwidth]{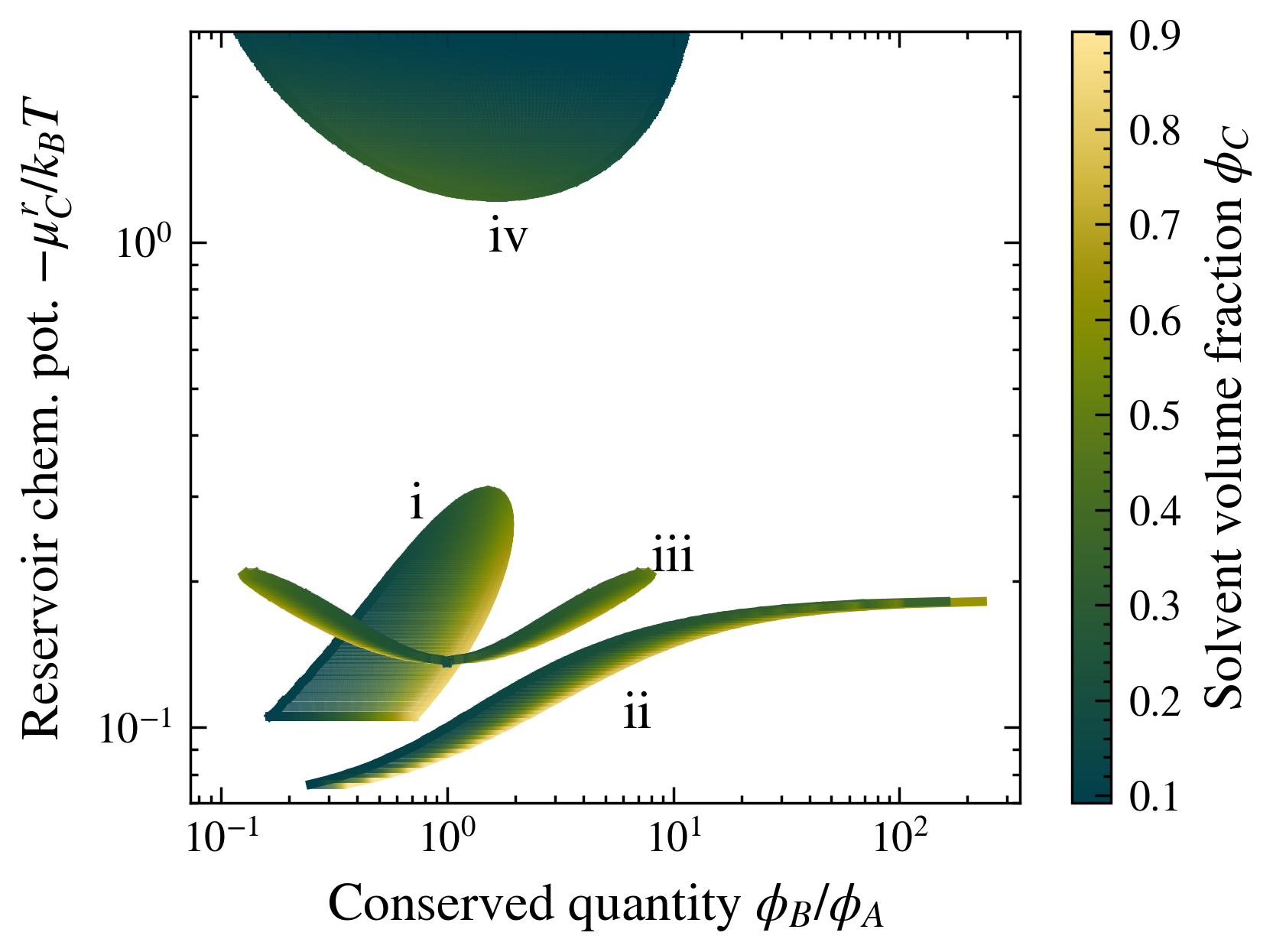}\label{fig:2c}}}
     \caption{\textbf{Evaporation/condensation phase diagrams:} (a) Evaporation/condensation equilibrium for different reservoirs are displayed together with the constraint line of constant $\phi_B/\phi_A$. The binodal encloses the blue-shaded area of phase coexistence, where the system phase separates into two phases connected by straight black tie lines. (b) For four sets of interaction strengths, creating four qualitatively different phase diagrams (i-iv), the orange-shaded region displays the initial states resulting in a phase-separated equilibrium. (c) The constraint of constant $\phi_B/\phi_A$ at phase-separated equilibria are qualitatively different for the different types of diagrams when varying the reservoir through $\mu_C^r$.}
     \label{fig:2}
\end{figure*}

To calculate the chemical potential, we consider the mean field Gibbs free energy $G$ per volume $V$ \cite{adame-arana_liquid_2020}:
\begin{align}
    \frac{G(T, p, \{\phi_k\})}{V} = k_\text{B}T &\sum_{i=0}^M \left[\frac{\phi_i}{\nu_i}\log{\phi_i} + \frac{\omega_i\phi_i}{\nu_ik_BT}\right] \nonumber\\ + \frac{1}{2} &\sum_{i=0}^M\sum_{j=0}^M \chi_{ij}\phi_i\phi_j+ p \, ,
\end{align}
where $\chi_{ij}$ is the interaction parameter characterizing the interaction between component $i$ and $j$. Moreover, $\omega_i$ and $\nu_i$ denote the internal energy and molecular volume of component $i$, respectively. The chemical potential of component $i$,
$\mu_i =  \partial{G}/{\partial N_i} |_{T,p,N_{j\neq i}}$,
thus becomes
\begin{align}
   \label{eq:def_mui}
    \mu_i &=   k_BT\left(\log{\phi_i} + 1\right) + \omega_i + \nu_i\left(p-\Lambda\right) + \nu_i \sum_{j=0}^M\chi_{ij}\phi_j \, , 
\end{align}
where $\Lambda$ is given by
\begin{equation}
    \Lambda = \frac{1}{2}\sum_{i=0}^M\sum_{j=0}^M\chi_{ij}\phi_i\phi_j + k_BT\sum_{i=0}^M\frac{\phi_i}{\nu_i}. \label{eq:lambda}
\end{equation}
The volume fraction of a component is defined by the volume it occupies relative to the total system volume,
$\phi_i = {\nu_i N_i}/{V}$, $ i \in \{1, 2, ..., M\}$. The solvent volume fraction $\phi_M$ is set by all other volume fractions
\begin{equation}
    \phi_M = 1 - \sum_{i=0}^{M-1}\phi_i \, .
\end{equation}
The total volume change due to evaporation and condensation is equal to the volume of molecules removed from or added to the system per unit of time
\begin{equation}
    \dot{V} = \sum_{i=0}^M\nu_i h_i\, . 
\end{equation}
Moreover, the volume change alters the volume fractions of all components, even components where $h_i=0$, as $\partial_t\phi_i=\nu_ih_i/V-\phi_i\dot{V}/V$, and thus affect their chemical potentials and chemical reactions. A reaction $\alpha$ is written as 
\begin{equation}
    \sum_{i=0}^M C_i\sigma_{i\alpha}^\rightharpoonup \rightleftharpoons \sum_{i=0}^M C_i\sigma_{i\alpha}^\leftharpoondown \, ,
\end{equation}
where $C_i$ denoted a chemical component and $\sigma_{i\alpha}^\rightleftharpoons$ are the are stoichiometric matrix elements~\cite{bauermann_chemical_2022}.
For the forward and backward chemical reactions rates $r_{\alpha}^\rightleftharpoons$~\cite{weber_physics_2019} we choose a symmetric form:
\begin{align}
    r_{\alpha}^\rightleftharpoons  &= k_{c,\alpha}\exp{\frac{\sum_{i=0}^M \sigma_{i\alpha}^\rightleftharpoons\mu_i}{k_BT}} \, .\label{eq:r_alpha}
\end{align}
From the chemical potential~\eqref{eq:def_mui}, we see that the forward and backward reaction rates are proportional to the product of their volume fractions $\phi_i$. This implies that once a component gets depleted, reactions slow down and arrest when reaching zero volume fractions. The ratio between the forward and backward reaction rates satisfies detailed balance of the rates~\cite{julicher_modeling_1997}:
\begin{equation}
     \frac{r_\alpha^\rightharpoonup}{r_\alpha^\leftharpoondown} = \exp{-\frac{\sum_{i=0}^M\sigma_{i\alpha}\mu_i}{k_BT}},
 \end{equation}
where we have defined $\sigma_{i\alpha}\equiv\sigma_{i\alpha}^\leftharpoondown -\sigma_{i\alpha}^\rightharpoonup$. 
The difference between the forward and backward reaction pathways, for $R$ different reactions, drives the net reaction rate
\begin{equation}
     r_i = \sum_{\alpha=1}^R \sigma_{i\alpha}\left(r_\alpha^\rightharpoonup - r_\alpha^\leftharpoondown\right). \label{eq:chem_react}
\end{equation}
If the interaction strengths cross a certain threshold value, the system will phase separate into two phases (denoted $\RN{1}$ and $\RN{2})$ with different composition $\phi_i^{\RN{1}/\RN{2}}$, but same chemical potential 
\begin{equation}
\mu_i^\RN{1} = \mu_i^\RN{2}. \label{eq:phase_coexistence}    
\end{equation}
The time evolution of the non-solvent volume fraction of a phase-separated system follows \cite{bauermann_chemical_2022,weber_physics_2019,bartolucci_selection_2022,bartolucci_controlling_2021}:
\begin{equation}
    \partial_t\phi_i^{\RN{1}/\RN{2}} = r_i^{\RN{1}/\RN{2}} - j_i^{\RN{1}/\RN{2}} + \frac{\nu_ih_i^{\RN{1}/\RN{2}}}{V^{\RN{1}/\RN{2}}} - \phi_i^{\RN{1}/\RN{2}}\frac{\dot{V}^{\RN{1}/\RN{2}}}{V^{\RN{1}/\RN{2}}} \, , \label{eq:kinetics}
\end{equation}
where $j_i^{\RN{1}/\RN{2}}$ gives the diffusive flux between the phases. The volume change of each phase for volume-conserving chemical reactions is given by
\begin{equation} \frac{\dot{V}^{\RN{1}/\RN{2}}}{V^{\RN{1}/\RN{2}}} = \sum_{i=0}^M \left(\frac{\nu_i h_i^{\RN{1}/\RN{2}}}{V^{\RN{1}/\RN{2}}} - j_i^{\RN{1}/\RN{2}}\right).
\end{equation}
To determine the diffusive flux $j_i$, we use  the $(M+1)$ constraints from phase equilibria (Eq. \eqref{eq:phase_coexistence}). See appendix~\ref{sec:PS_kin} for further discussion. For a homogeneous system, the governing equations are retrieved by setting $j_i^{\RN{1}/\RN{2}}=0$, and dropping the phase specifying superscript. Accounting for transition states in the theory leads to the same kinetic equations, as long as the transition state is short-lived; see appendix~\ref{sec:transition_states} for a detailed discussion.

For a phase-separated system, the chemical reactions concomitantly occur in both phases, while the fluxes related to evaporation/condensation are considered to occur at a surface area enclosing the dilute phase. As the two phases are at phase equilibrium, the choice of which phase is in contact with the reservoir becomes irrelevant. Furthermore, we will assume that the reaction rate coefficient does not depend on the phase $k_{c,\alpha}^\RN{1}=k_{c,\alpha}^\RN{2}$. We consider that evaporation/condensation of macromolecules is negligible compared to the solvent~\cite{loche_molecular_2022}, such that $k_{e,i}=k_e\delta_{iC}$. When only the solvent is evaporating/condensing, and no chemical reactions occur, the ratio of each non-solvent volume fraction, $\phi_i/\phi_j$, is conserved. To reduce dimensionality and the number of free parameters, we will consider a single mono-molecular reaction $A\rightleftharpoons B$, such that $\sigma_{i\alpha}^\rightharpoonup = \delta_{iA}\delta_{\alpha1}$, $\sigma_{i\alpha}^\leftharpoondown = \delta_{iB}\delta_{\alpha1}$, and $k_{c,\alpha}=k_c\delta_{\alpha,1}$. The kinetic algorithm to evolve the system according to Eq.~\eqref{eq:kinetics} builds on a method derived by Bauermann and Laha~\cite{bauermann_chemical_2022}; see Appendix~\ref{sec:PS_kin} for how the theory accounts for the phase equilibrium constraint, and Appendix~\ref{seq:limits} for the properties a system should have for phase equilibrium to hold well at each time during the kinetics.

\section*{Results}

\subsection*{Thermodynamics of evaporation/condensation mixtures without chemical reactions}

\begin{figure*}
    \centering
    \makebox[\textwidth][c]{
     \sidesubfloat[]{\includegraphics[height=0.28\textwidth]{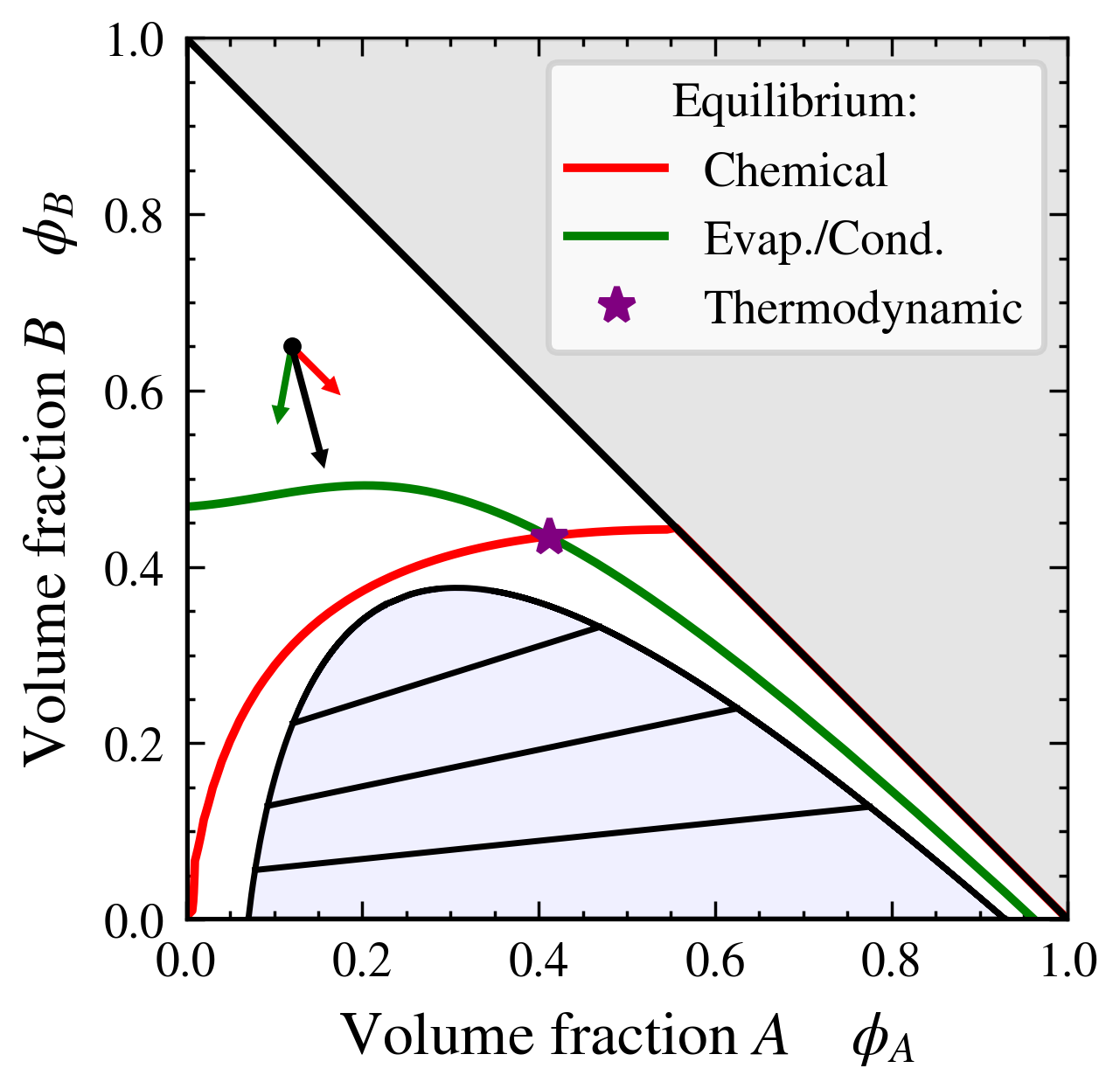}\label{fig:3a}}
     \sidesubfloat[]{\includegraphics[height=0.28\textwidth]{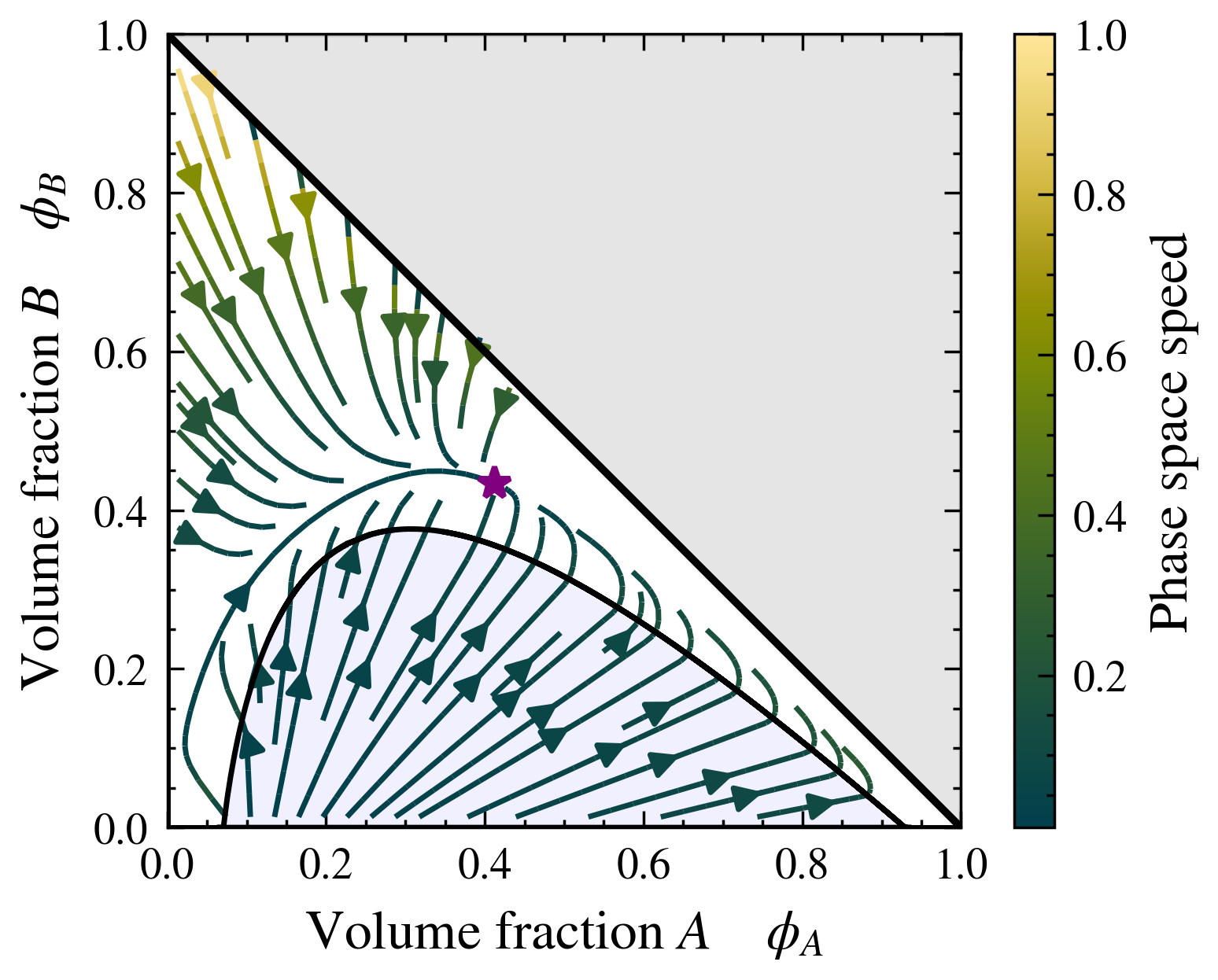}\label{fig:3b}}
     \sidesubfloat[]{\includegraphics[height=0.28\textwidth]{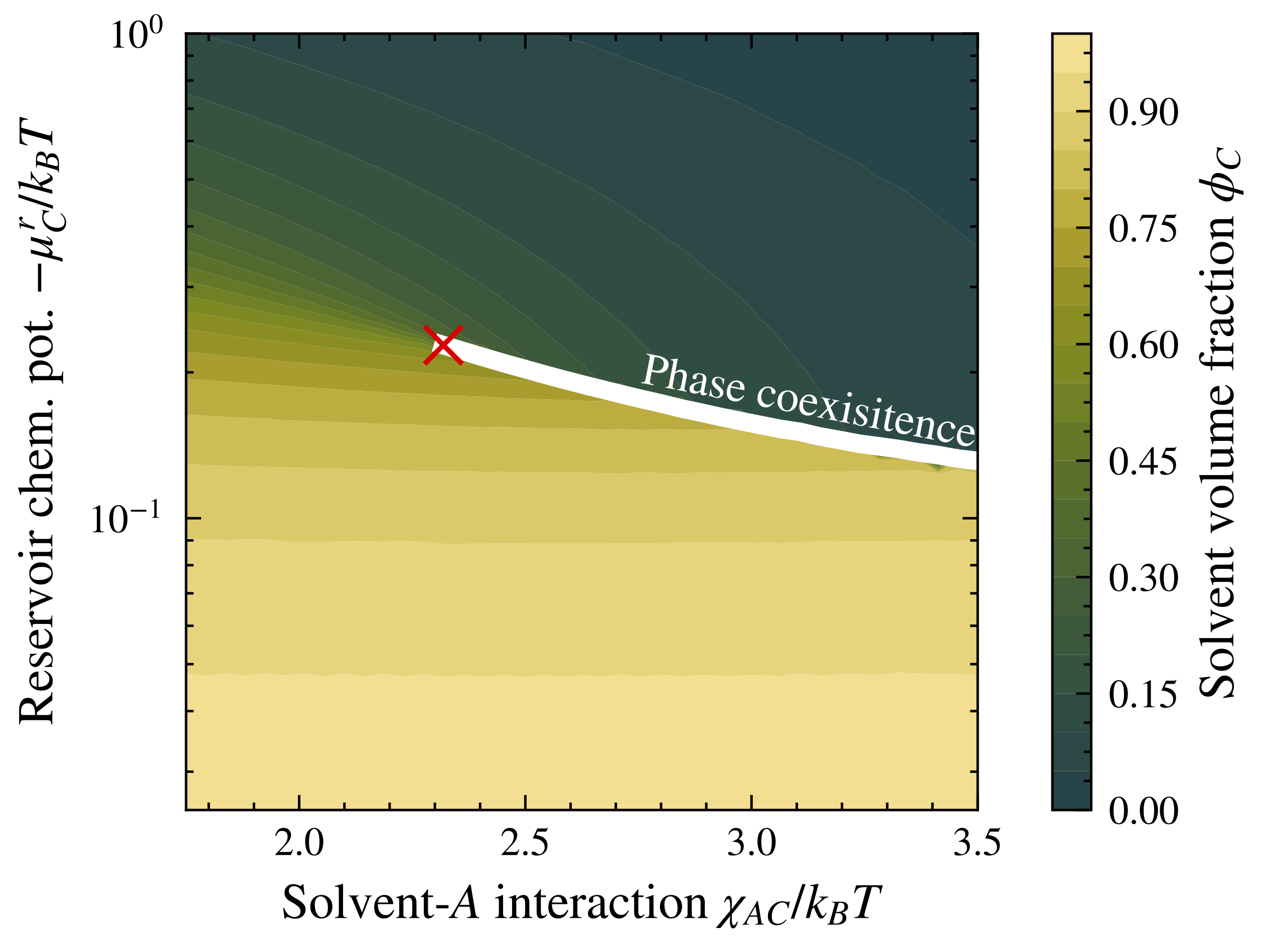}\label{fig:3c}}}
     \caption{
     \textbf{Evaporation/condensation and chemically reacting phase diagrams:} (a) The intersection of the equilibrium lines of evaporation/condensation and the chemical reaction gives the thermodynamic equilibrium point all initial states converge towards. An example of the change in volume fraction is displayed as a black arrow, which is the sum of a chemical (red) and evaporation/condensation component (green). (b) The set of all such points produces a flow field in the phase diagram, all converging at the thermodynamic equilibrium point, displayed for $k_e/k_c=4$. (c) Phase-separated thermodynamic equilibrium states are only achievable for a single reservoir value for each interaction strength. The phase coexistence line separates solvent-poor and solvent-rich states until the critical point (red cross).}
     \label{fig:3}
\end{figure*}

To understand chemical kinetics in systems that can phase separate and undergo wet-dry cycles, 
it is instrumental first to understand equilibrium states in the absence of chemical reactions.
Equilibrium states composed of coexisting phases form through evaporation/condensation if the equilibria are compatible. Compatible equilibria refer to the case where the phase equilibrium condition~\eqref{eq:phase_coexistence} is concomitantly satisfied together with one or more other equilibrium conditions~\cite{bauermann_chemical_2022}. In absence of chemical processes,  this other condition corresponds to evaporation-condensation equilibrium.
Graphically, such compatible equilibria correspond to the intersections in Fig.~\ref{fig:2a} between the binodal and the mononodal line that is set by the reservoir chemical potential $\mu_C^\text{r}$ (orange dashed line).
During evaporation and condensation $\phi_B/\phi_A$ is conserved.
Thus, the domain of compatible equilibria  is cone-shaped and
is spanned by two distinct conservation lines of constant $\phi_B/\phi_A$ (orange domains in Fig.~\ref{fig:2b}).
Any initial state will move during its evaporation/condensation kinetics and settle in a compatible equilibrium with coexisting phases. 
The number and the area of domain with compatible equilibria depend on the interactions among the molecules.
The area of compatible domains increases when $A$ and $B$ interact more differently with the solvent (case i in contrast to ii and iii), or when $A$ and $B$ phase separate independently of the solvent (case iv). 

\begin{figure*}
     \makebox[\textwidth][c]{
      \sidesubfloat[]{\includegraphics[height=0.25\textwidth]{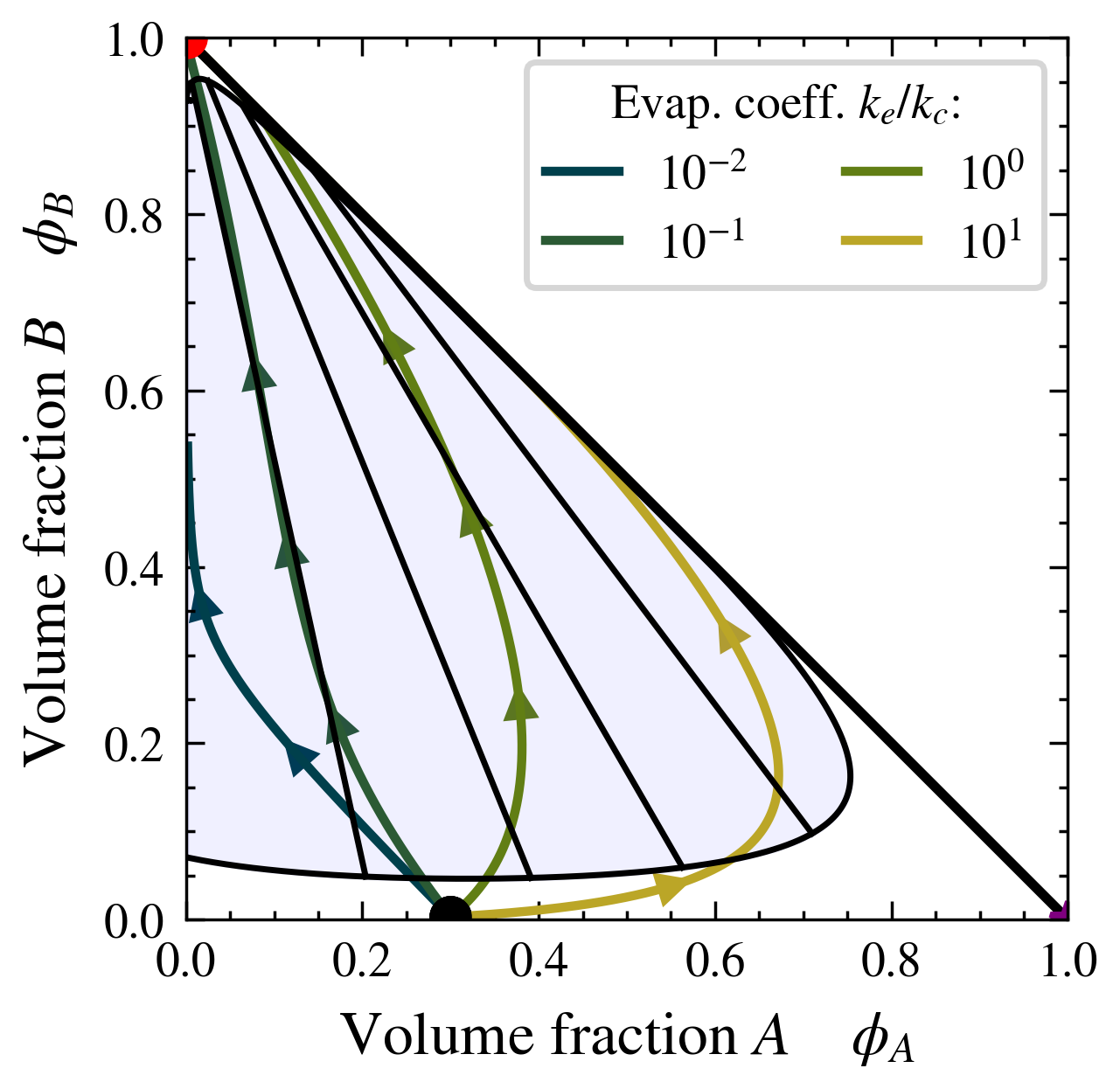}\label{fig:4a}}
      \sidesubfloat[]{\includegraphics[height=0.25\textwidth]{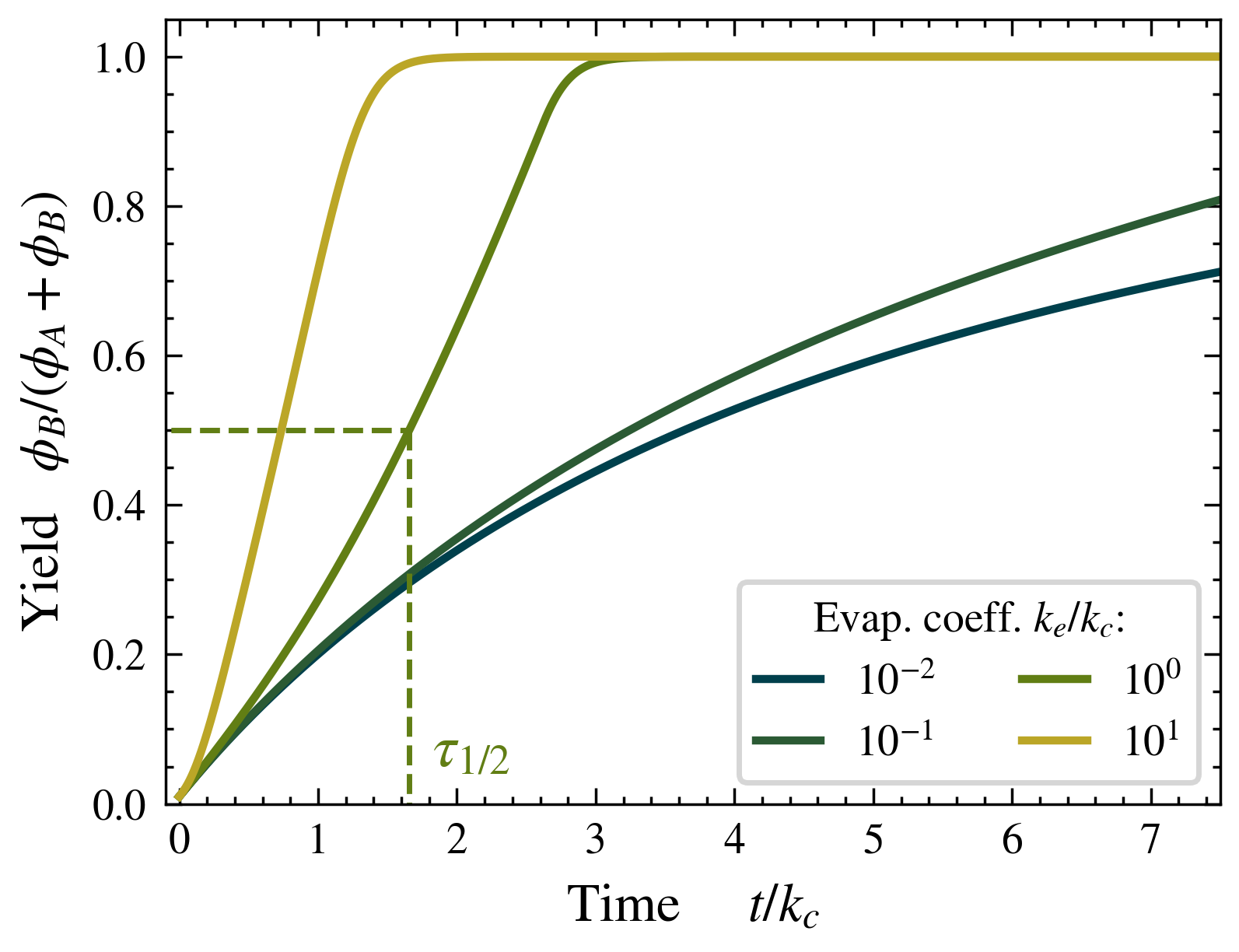}\label{fig:4b}}
      \sidesubfloat[]{\includegraphics[height=0.25\textwidth]{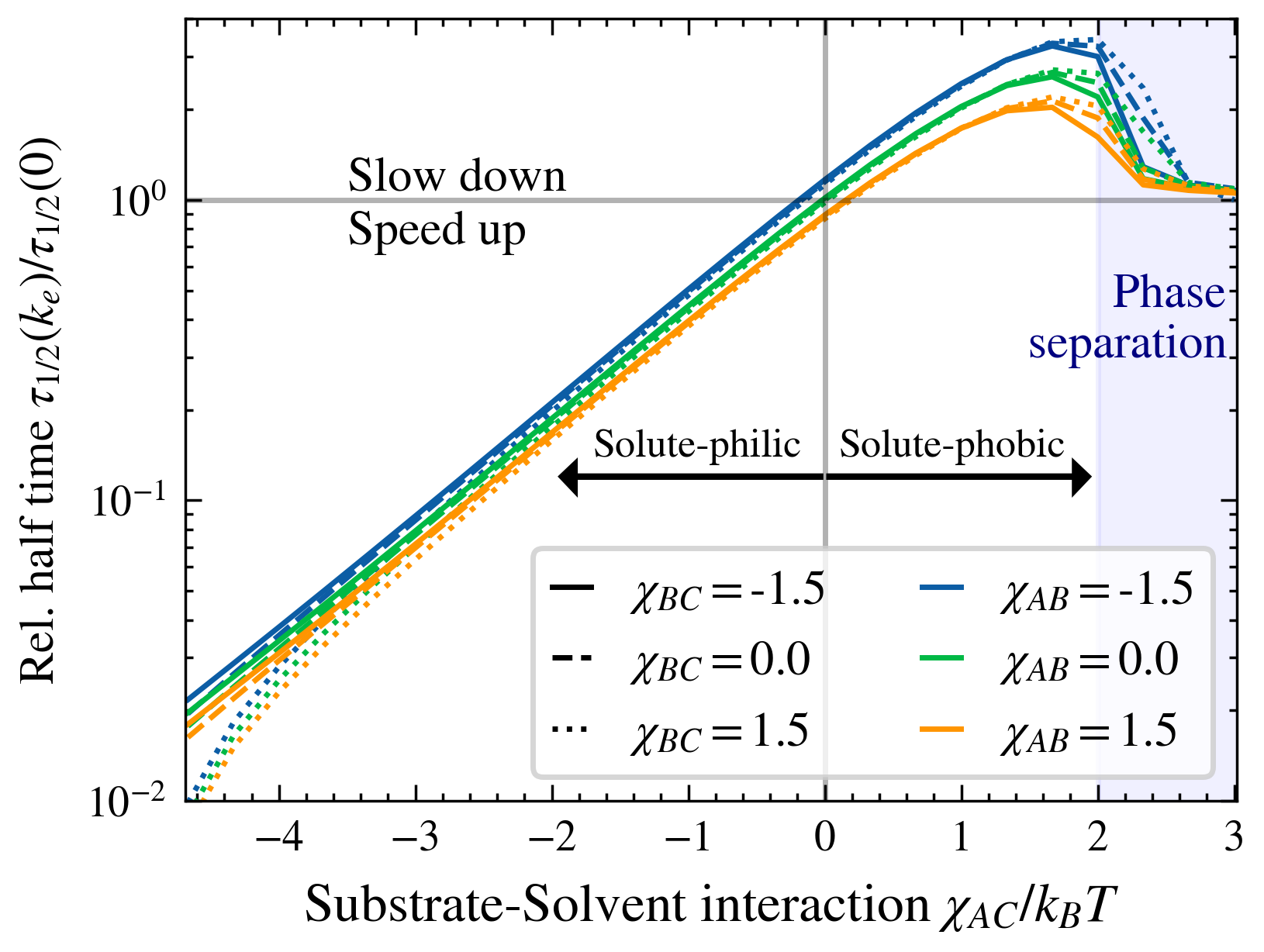}\label{fig:4c}}}\\
     \makebox[\textwidth][c]{
      \sidesubfloat[]{\includegraphics[height=0.25\textwidth]{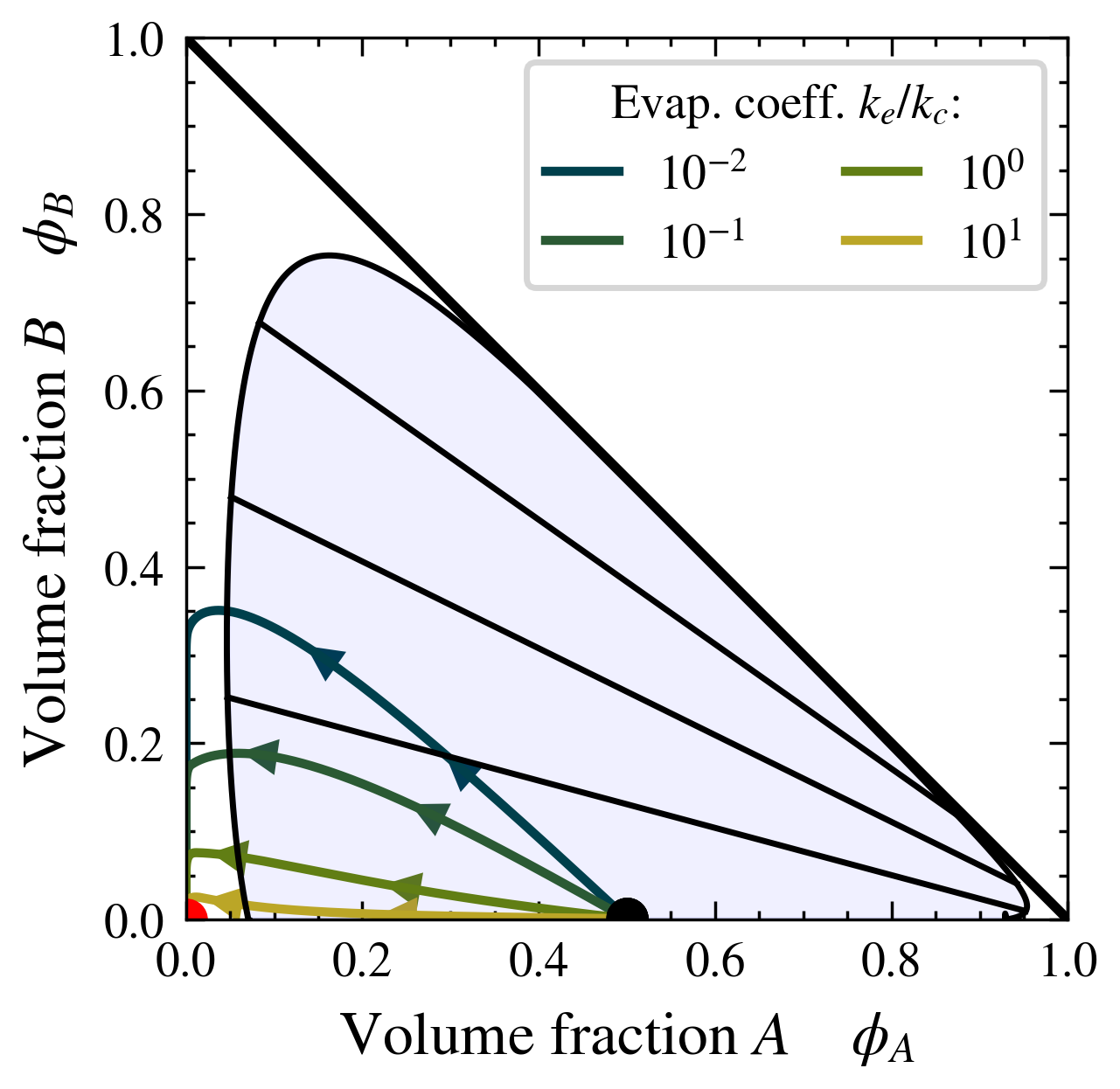}\label{fig:4d}}
      \sidesubfloat[]{\includegraphics[height=0.25\textwidth]{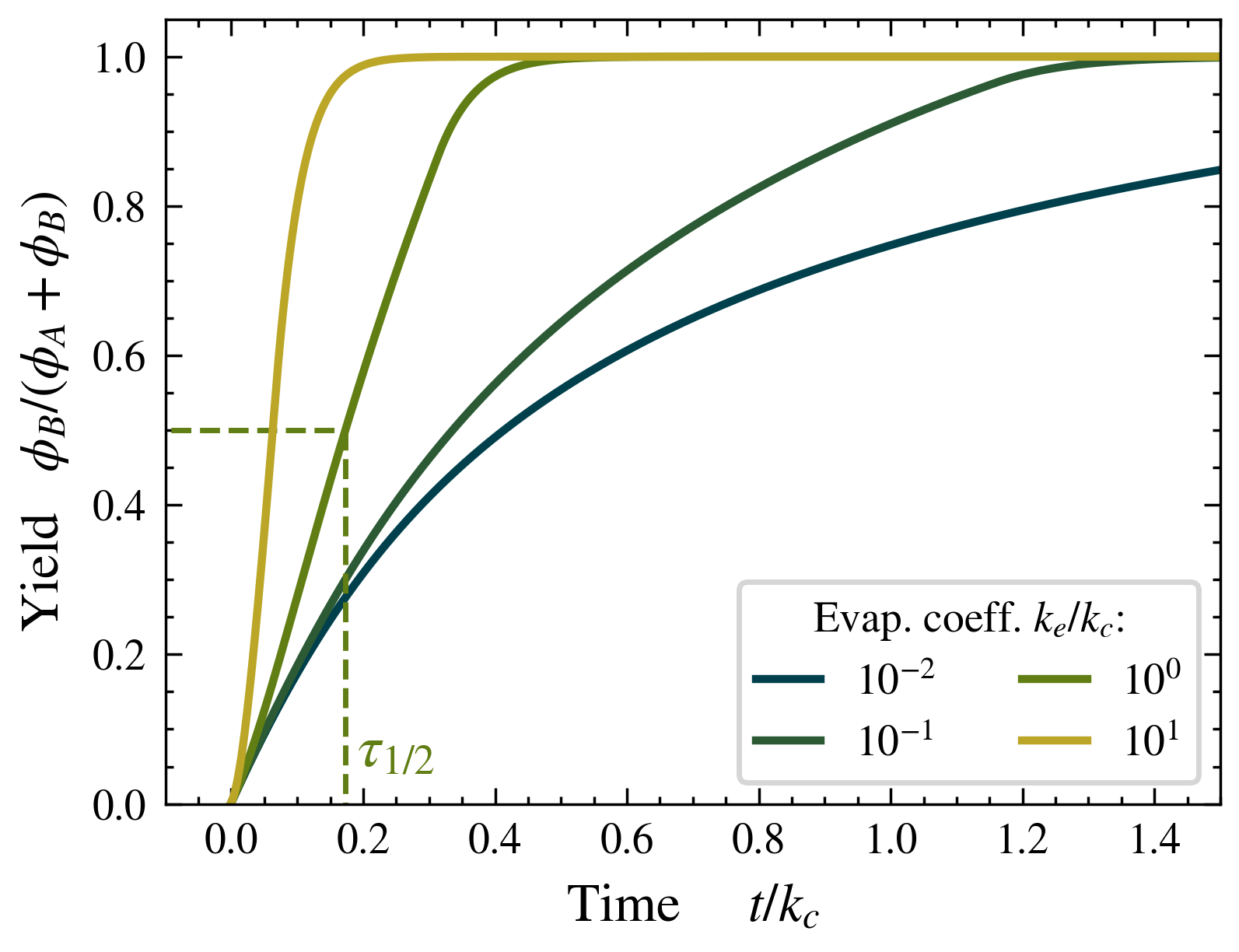}\label{fig:4e}}
      \sidesubfloat[]{\includegraphics[height=0.25\textwidth]{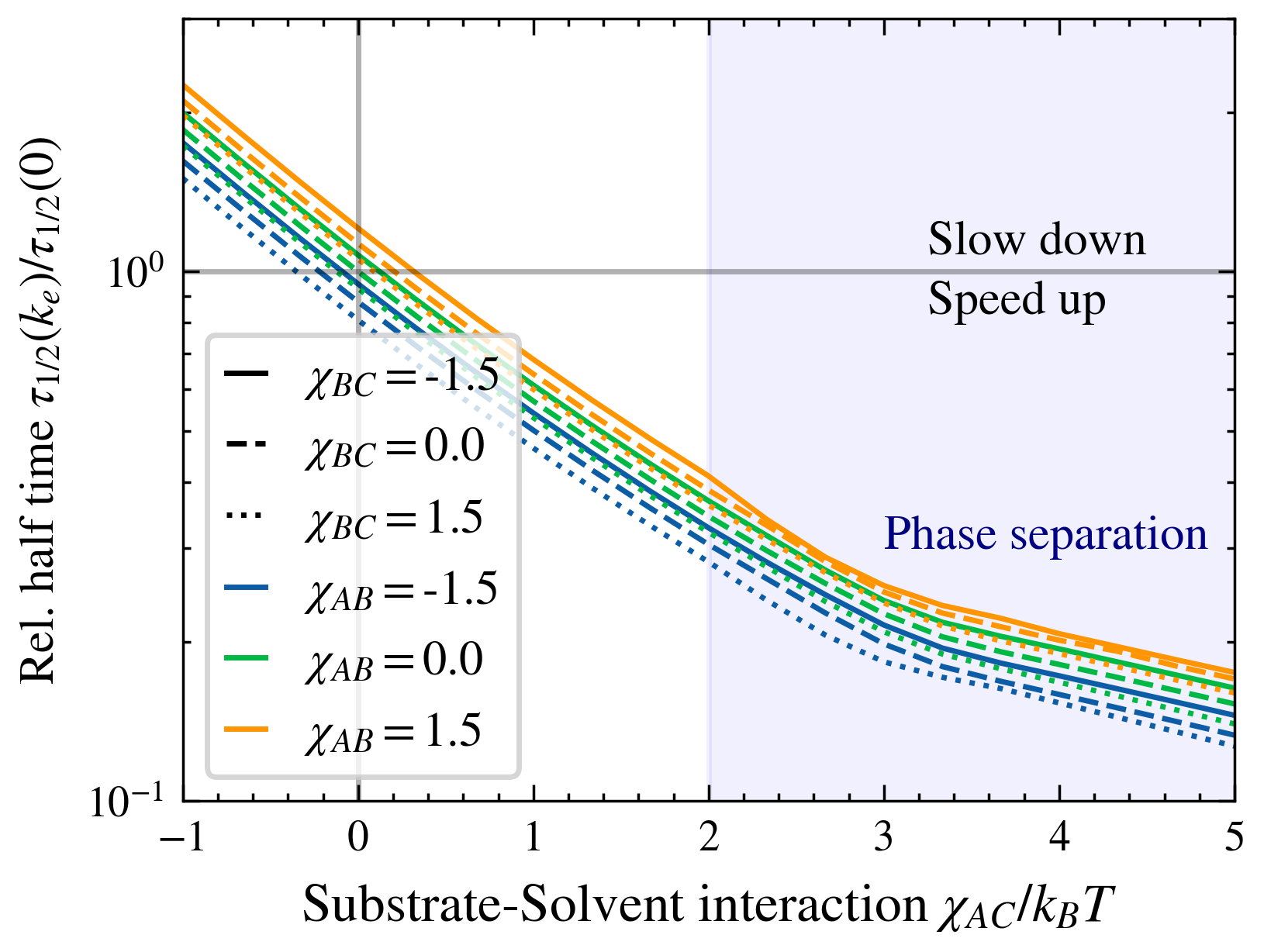}\label{fig:4f}}}
     \caption{\textbf{Chemical speed-up through evaporation/condensation:}
     The pathway through the phase diagram during evaporation (a-c) and condensation (d-f) towards the same equilibrium (red dot) for different $k_e/k_c$ are shown in (a) and (d), respectively. The yield of the product $B$, $\phi_B/(\phi_A+\phi_B)$, is displayed as a function of time in (b) and (e) showing a significant speed-up through the different pathways. How this speed-up depends on the interaction strengths for $k_e/k_c=10$ is displayed in (c) and (f). The top row (a-c) shows the results for evaporation, and the bottom row from condensation (d-f).
     }\label{fig:4}
\end{figure*}

The equilibrium states are homogeneous when evaporation-condensation and phase equilibrium are incompatible. Such incompatible states graphically correspond to the complement of the orange shaded domains in Fig.~\ref{fig:2b}, i.e., the values of $\phi_B/\phi_A$ that lie outside of the binodal.
Incompatible equilibria are favoured when the $A$ and $B$ interact similarly with the solvent, 
and when neither $A$ or $B$ phase separates with the solvent, as discussed for compatible equilibria. All equilibria are incompatible for some values of the reservoir chemical potential (green and yellow lines in Fig.~\ref{fig:2a}).

The phase behavior of the evaporating/condensing mixtures (i-iv) without chemical reactions can be summarized by the thermodynamic phase diagram spanned by the reservoir chemical potential $\mu_C^\text{r}$ and the conserved variable $\phi_B/\phi_A$. 
The different interactions (i-iv) lead to very different shapes where two phases can coexist with very different values of the conserved variable $\phi_B/\phi_A$,
as shown in  Fig.~\ref{fig:2c}. For case iv, where the coexisting phases have a similar solvent volume fraction, the non-solvent components phase separates independently of the solvent. As a result, each phase has a large difference in $\phi_B/\phi_A$, creating a significant phase-separated domain.

\subsection*{Thermodynamics of evaporation/condensation mixtures with chemical reactions}

In the presence of chemical reactions, $\phi_B/\phi_A$ is not conserved, and compatible equilibria are no longer achievable.
Thermodynamic equilibrium, therefore, always corresponds to a homogeneous state where the conditions of chemical equilibrium ($\mu_A=\mu_B$) and evaporation-condensation equilibrium ($\mu_C=\mu_C^r$) are fulfilled; see the intersection between the red and green line in Fig.~\ref{fig:3a}.
Any initial state $(\phi_A,\phi_B)$ will evolve following a flow field in the phase diagram toward the fixed point of thermodynamic equilibrium; see Fig.~\ref{fig:3b}.
Each point in this flow diagram has two independent directions that characterize the rates of change in the average volume fractions, which correspond to  
the chemical turnover flux of constant $\phi_A+\phi_B$ and evaporation-condensation flux of constant $\phi_B/\phi_A$; see vectors shown in Fig.~\ref{fig:3a}. 
Such fluxes of strength visualized by the length of each vector are set by $k_e$ and $k_c$. 
Gibbs' phase rule (discussed in appendix~\ref{sec:Gibbs}) makes the domain of coexisting states inaccessible at thermodynamic equilibrium, such that small changes in reservoir chemical potential $\mu_C^\text{r}$ close to the binodal line pronounces significant changes in the equilibrium compositions. This behavior is shown in Fig.~\ref{fig:3c}, where solvent-rich and solvent-poor equilibrium states are separated by the phase coexistence line ending up at the critical point (red cross) in Fig.~\ref{fig:3c}.

\begin{figure*}
    \centering
     \makebox[\textwidth][c]{
      \sidesubfloat[]{\includegraphics[height=0.27\textwidth]{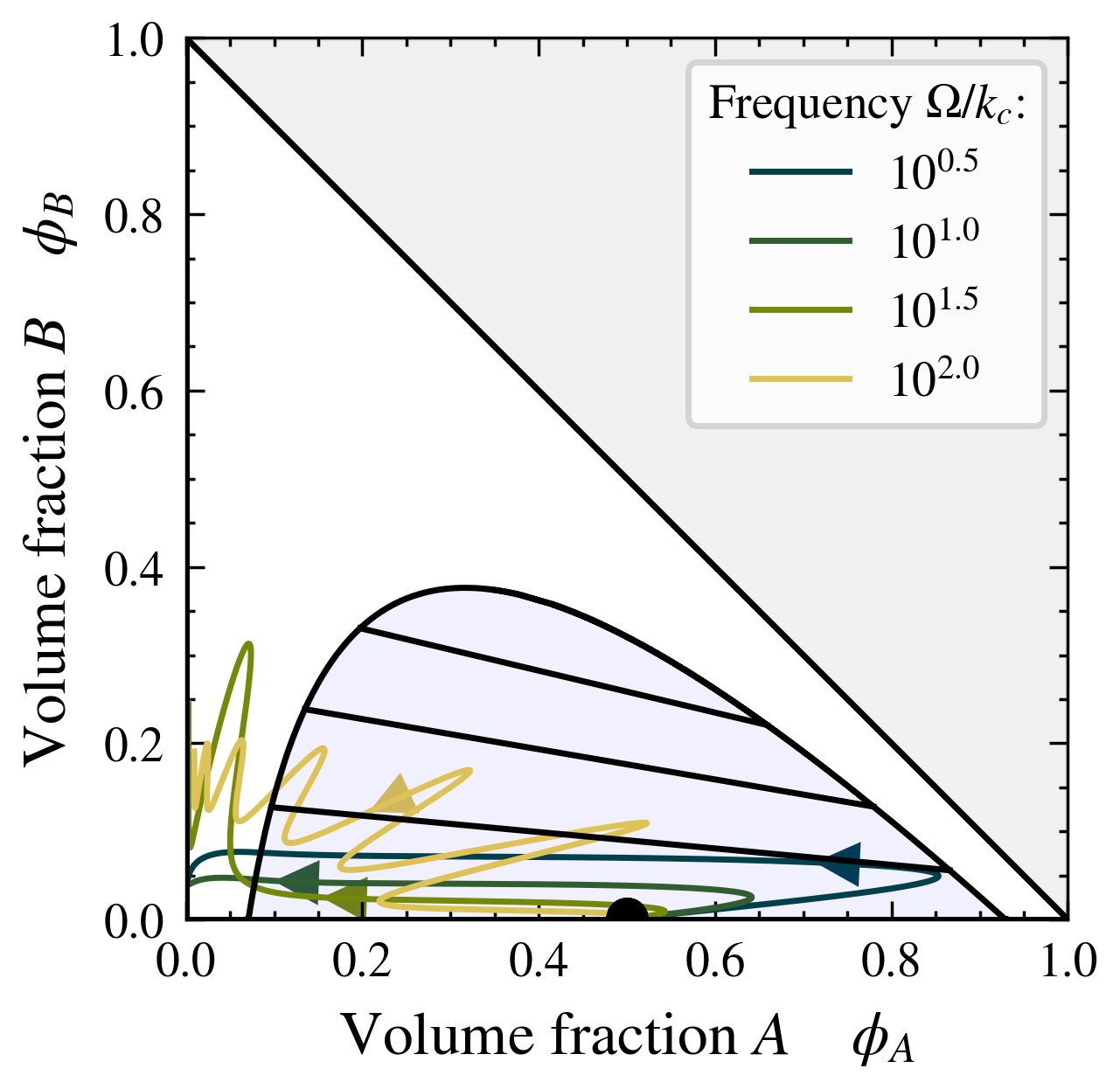}\label{fig:5a}}
      \sidesubfloat[]{\includegraphics[height=0.27\textwidth]{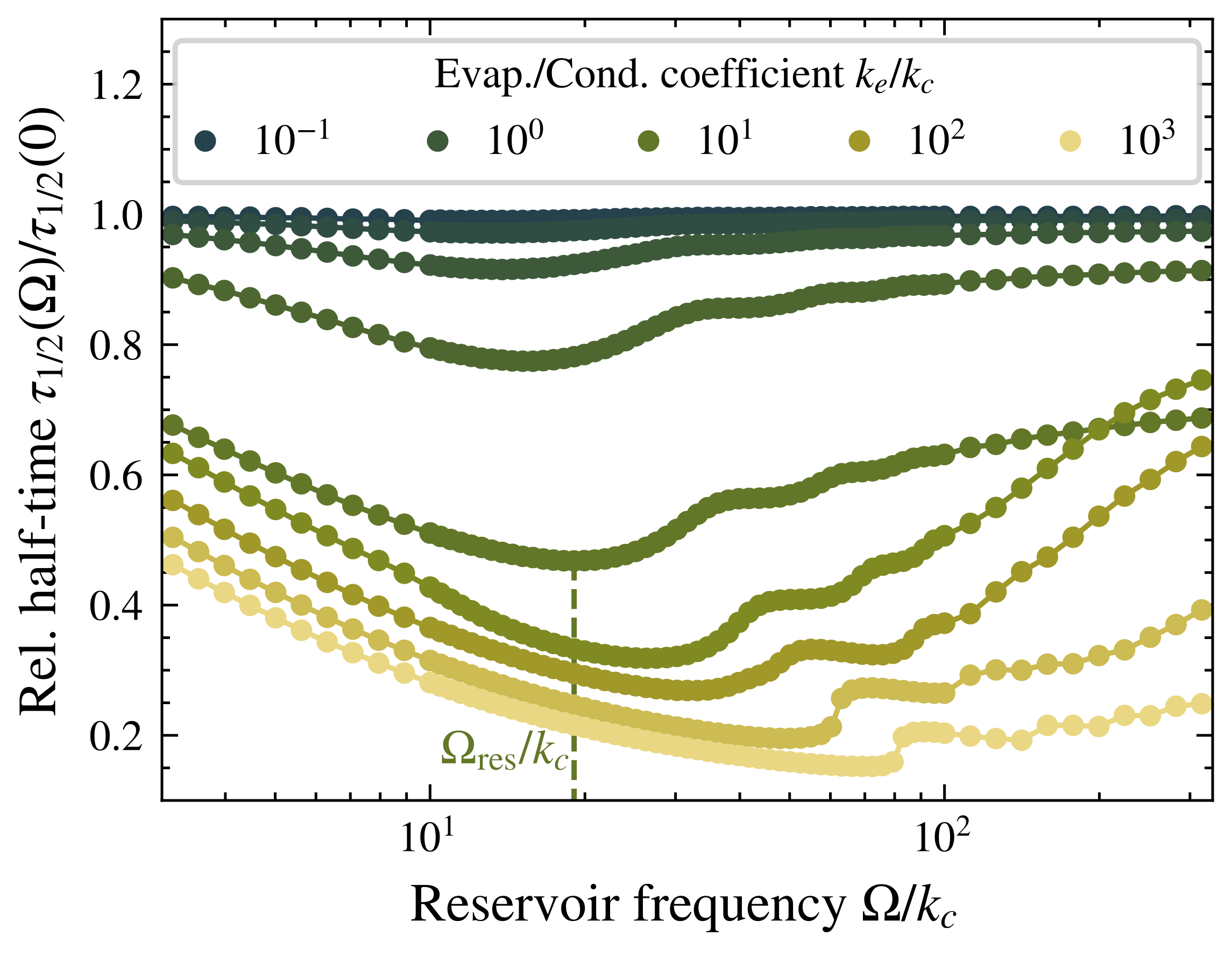}\label{fig:5b}}
      \sidesubfloat[]{\includegraphics[height=0.27\textwidth]{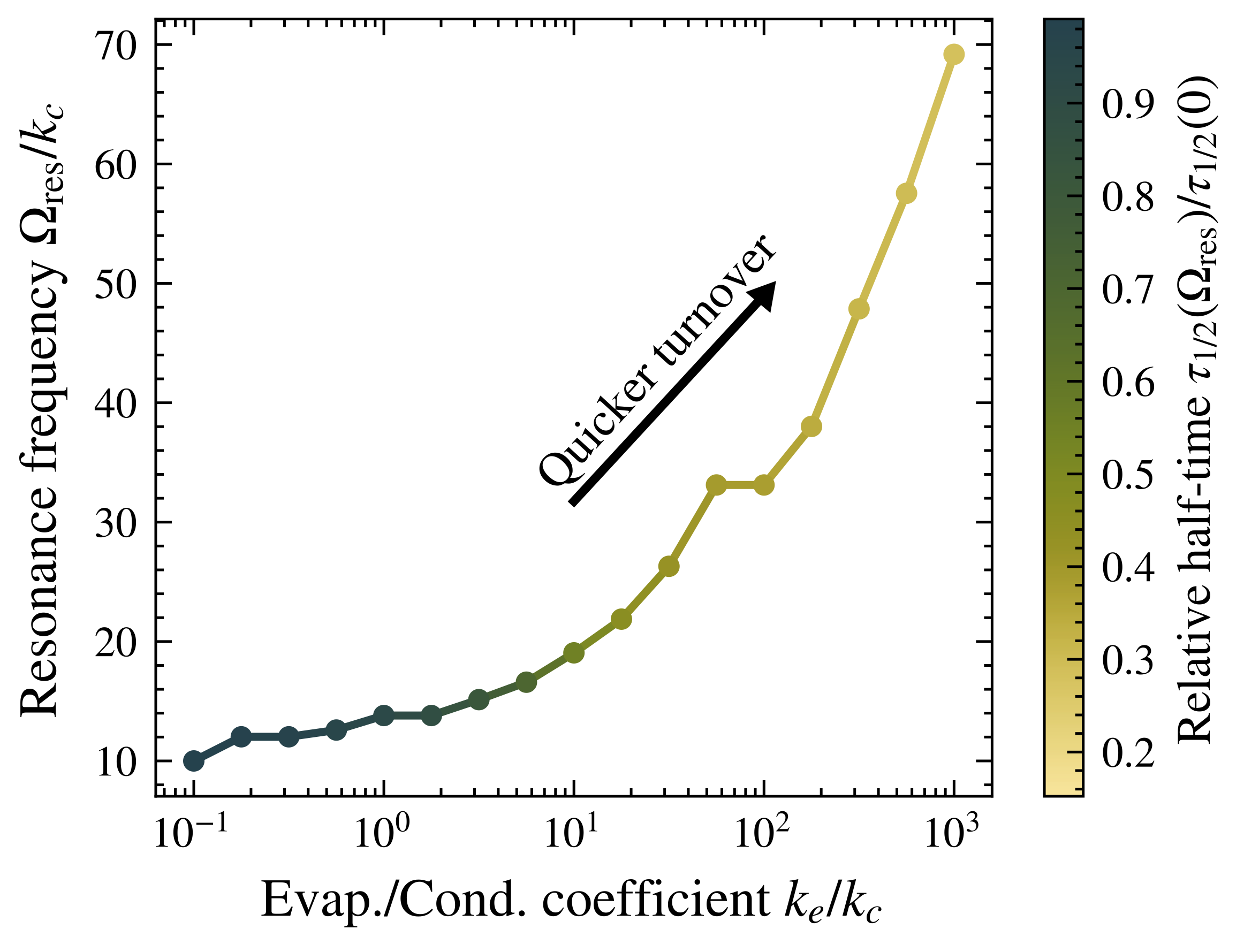}\label{fig:5c}}}\\
     \makebox[\textwidth][c]{
      \sidesubfloat[]{\includegraphics[height=0.27\textwidth]{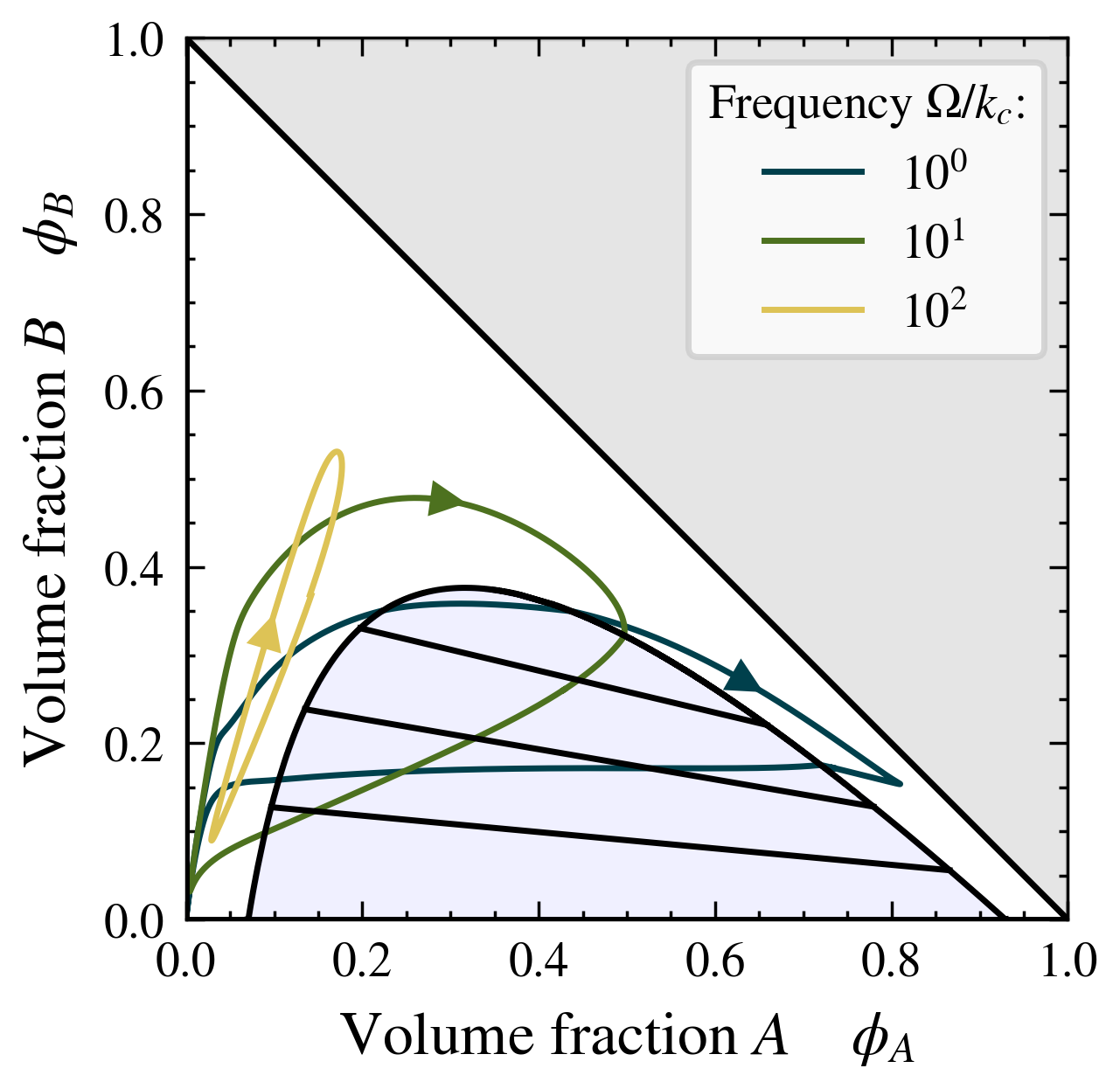}\label{fig:5d}}
      \sidesubfloat[]{\includegraphics[height=0.27\textwidth]{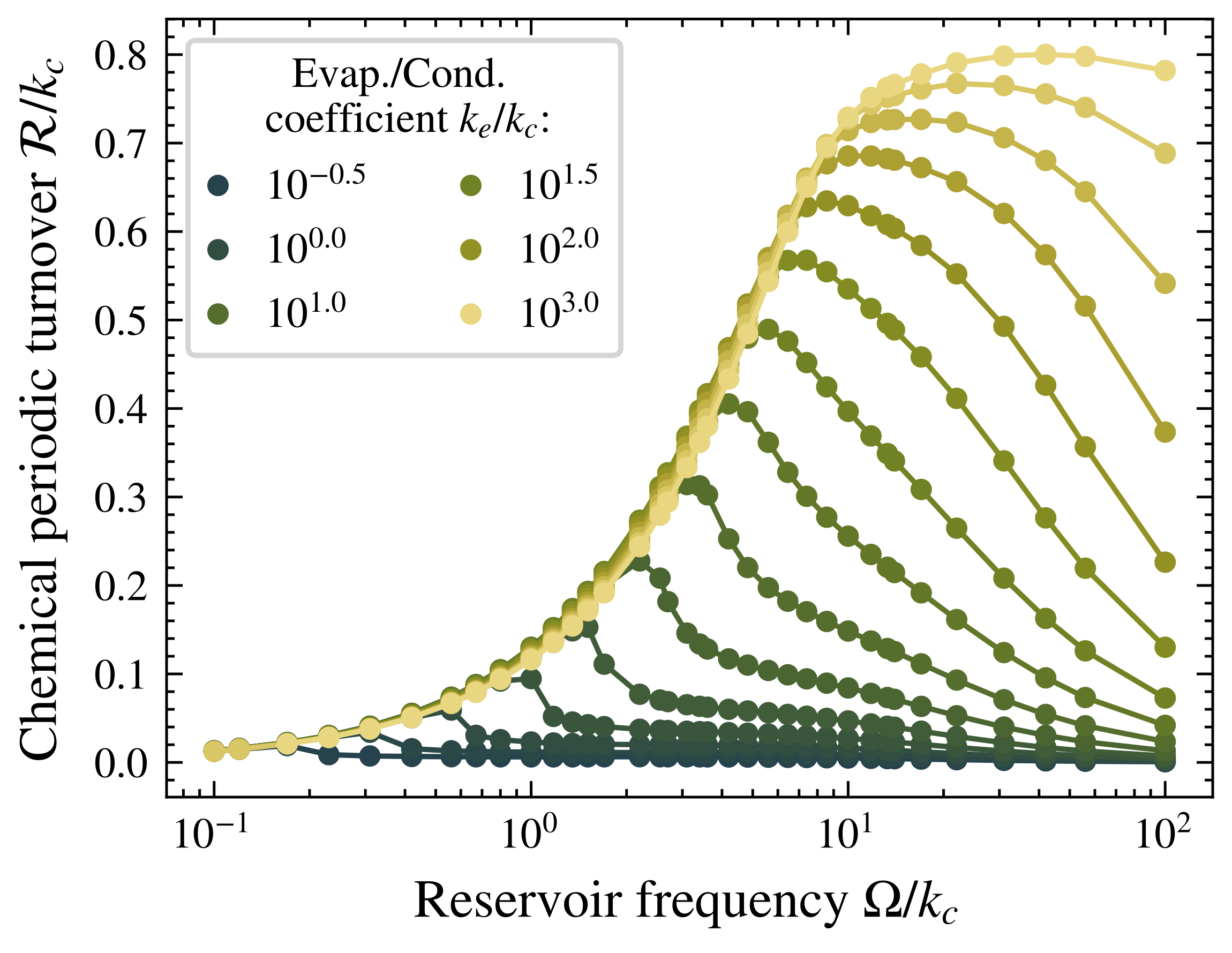}\label{fig:5e}}
      \sidesubfloat[]{\includegraphics[height=0.27\textwidth]{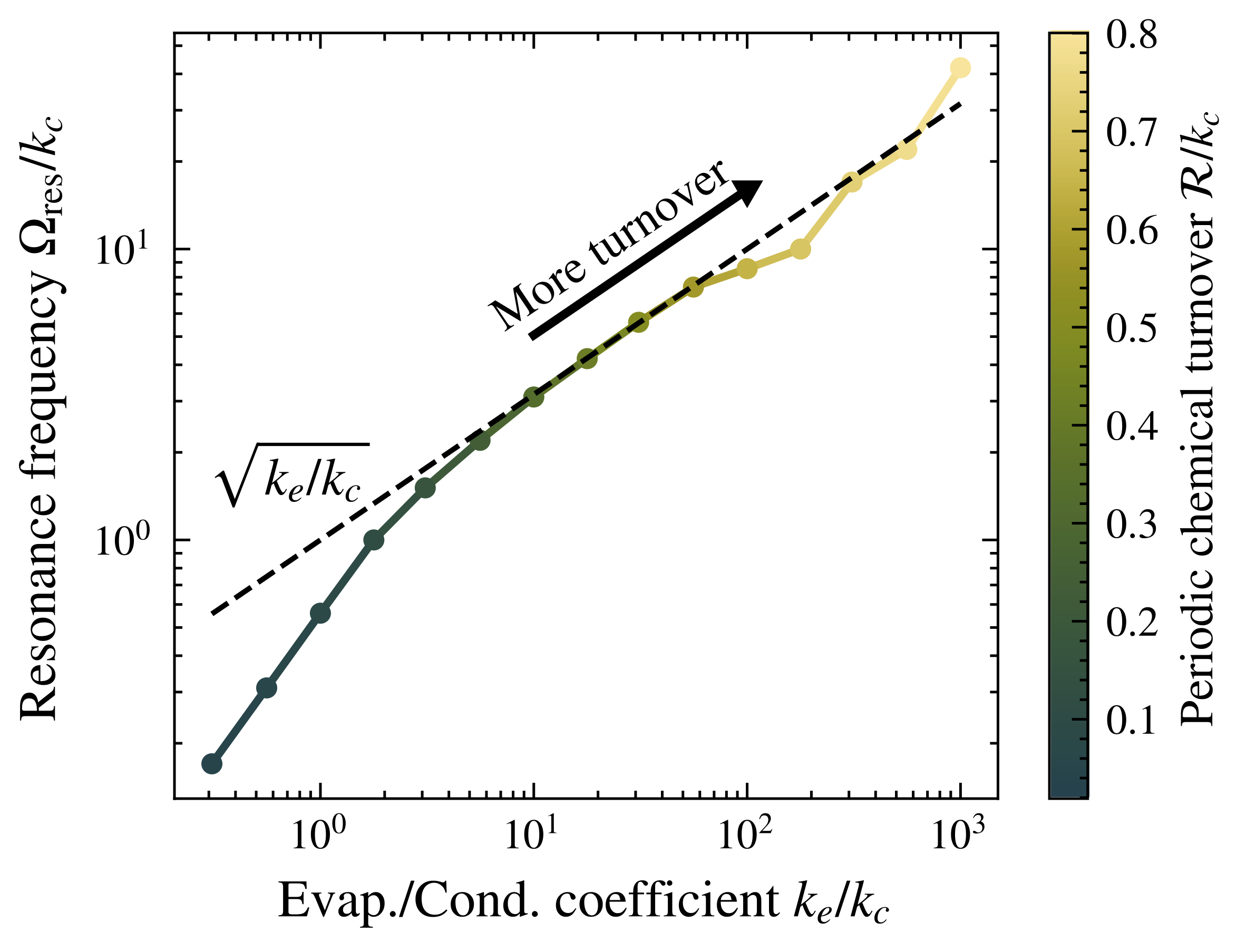}\label{fig:5f}}}
     \caption{\textbf{Resonance behavior in chemical reactions with an oscillating reservoir:}
     The top row (a-c) shows the conversion from $A$ to $B$ on the path towards an orbit, while the bottom row (d-f) shows equivalent quantities during the orbits. The change in the phase-diagram pathway with cycling frequency $\Omega$ is displayed in (a) and (d) for $k_e/k_c=10^{1.5}$ and $10^{2.75}$ respectively. Both cases exhibit a resonance behavior between the cycling frequency of the reservoir and the chemical turnover, quantified by the conversion half-time (b,c) and periodic chemical turnover (Eq. \eqref{eq:def_R}) (e,f). Both values increase with $k_e/k_c$ at the resonance frequency.}
     \label{fig:5}
\end{figure*}

\subsection*{Speed-up and slow-down of chemical reactions through evaporation and condensation}

Evaporation and condensation can speed up chemical reactions by driving the system along different phase diagram trajectories. 
For unidirectional chemical reactions, where $\omega_\text{product}\ll\omega_\text{substrate}$, these trajectories are characterized by solvent-poor or solvent-rich conditions during the relaxation to thermodynamic equilibrium (Fig.~\ref{fig:4}(a,d)). These different solvent conditions have a dramatic impact on the time $(\tau_{1/2})$ it takes to turn over $50\%$ of $A$ to $B$, seen as the half-time of the yield $\phi_B/(\phi_A+\phi_B)$ in Fig.~\ref{fig:4}(b,e). For the depicted molecular interactions, i.e., $A$ favors the solvent while $B$ does not, a quick removal of the solvent (large $k_e/k_c$) causes an increase in the chemical potential of $A$. This increase in the substrate's chemical potential, visualized in Fig.~\ref{fig:1a}, speeds up the turnover from $A$ to $B$.

The threshold between slow-down and speed-up of chemical reactions corresponds to no interactions between the components $(\chi_{ij}=0)$, as seen in Fig.~\ref{fig:4c} and \ref{fig:4f}. Thus, solely up-concentrating the components does not affect the speed-up, whose origin is a pure consequence of non-dilute conditions and the interactions among reacting components. The leading term to quantify the effect is the substrate-solvent interaction, which enters exponentially as $\chi_{AC}\phi_C$ in the reaction rate in Eq.~\eqref{eq:chi_speedup}. The sign of $\chi$ determines whether evaporation or condensation aids the chemical turnover, while the magnitude determines the speed-up. In other words, reactions of solvophilic substrates are sped up through evaporation, and solvophobic substrates under condensation. Though other interactions $(\chi_{AB}, \chi_{BS})$ also affect the speed-up, they are less relevant than the substrate-solvent interaction $(\chi_{AS})$, as seen in figure 4c and 4f. All statements about interaction strengths are qualitatively reversed once we switch from evaporation (Figs. \ref{fig:4}(a-c)) to condensation (Figs. \ref{fig:4}(d-f)).

\subsection*{Chemical reactions subject to wet-dry cycles}

Now we discuss the effects on chemical reactions when the system is subject to cycles  of the reservoir chemical potential
\begin{equation}
    \mu_C^r(t) = \langle \mu_C^r \rangle + \mu_{C, r}^{\text{amp}}\sin{\left(\Omega t\right)} \, .\label{eq:oscillations}
\end{equation}
Here, $\Omega$ denotes the cycling frequency, $\langle \mu_C^r \rangle$ the average reservoir value and $\mu_{C, r}^{\text{amp}}$ the oscillation amplitude.
We now discuss two cases of cycles: (i) the chemical path toward the orbit, and (ii) the continuous movement among this chemical orbit. For case (i), we are interested in the chemical turnover for a unidirectional chemical reaction, as before, and how it is affected by cycles. After the initial phase, the trajectory becomes periodic. In case (ii), we are interested in the chemical turnover rate during such an orbit.

(i) We find that wet-dry cycles can speed up chemical turnover relative to no cycling, with a resonance peak at a specific cycling frequency. Different cycling frequencies $\Omega$ result in the different pathways displayed in Fig.~\ref{fig:5a}, with different conversion half-times. Oscillating reservoirs decrease the conversion half-time relative to no oscillations, $\tau_{1/2}(\Omega)/\tau_{1/2}(0)$, by up to a factor of $5$ for the parameters used in Fig.~\ref{fig:5b}. The resonance cycling frequency $\Omega_{\text{res}}$ is stable, changing less than a decade when varying $k_e/k_c$ over four decades, displayed in Fig.~\ref{fig:5c}. Slow chemical reactions have the largest speed-up due to wet-dry cycles.

(ii) With repeated cycling, the system will eventually enter an oribt, where the path in the phase diagram is a closed loop (Fig.~\ref{fig:5d}). To quantify the chemical activity per period, we define the relative chemical turnover per cycle, 
\begin{equation}
     \mathcal{R} = \frac{\Omega}{2\pi}\int_0^{2\pi/\Omega} \dif t \lvert r^{\rightharpoonup}-r^{\leftharpoondown} \rvert \, . \label{eq:def_R} 
 \end{equation}
This quantity is at its minimal $\mathcal{R}=0$ if no chemical reactions take place during the cycle, and maximal $\mathcal{R}=\Omega/\pi$ if all molecules undergo both the forward and backward reaction per cycle.

Different cycling frequencies $\Omega/k_c$ produce different orbits in the phase diagram, as shown in Fig.~\ref{fig:5d}. Small $\Omega/k_c$ allows the system to remain close to chemical equilibrium, while large values do not allow the system to respond to the change in the reservoir. In between, there exists a frequency where the system has time to react to the reservoir and be driven away from the chemical equilibrium line, maximizing $\mathcal{R}$. For larger values of $k_e/k_c$, the ability of the system to respond to the reservoir increases, making $\mathcal{R}$ larger as well. With increasing $k_e/k_c$, the system is able to follow the reservoir equilibrium, saturating $\mathcal{R}$. Without interactions $(\chi_{ij}=0)$, the chemical equilibrium $(\mu_A=\mu_B)$ is set by $\phi_B/\phi_A=\text{constant}$. As the equilibrium condition is identical to the evaporation/condensation constraint, reservoir oscillations cannot drive the system away from chemical equilibrium. Thus, again, changes in the solvent composition can only affect chemical reactions when non-dilute interactions and interactions among reacting components are considered.

\begin{figure}
    \begin{center}
     \makebox[0.5\textwidth][c]{
     \subfloat{\includegraphics[width=0.85\textwidth]{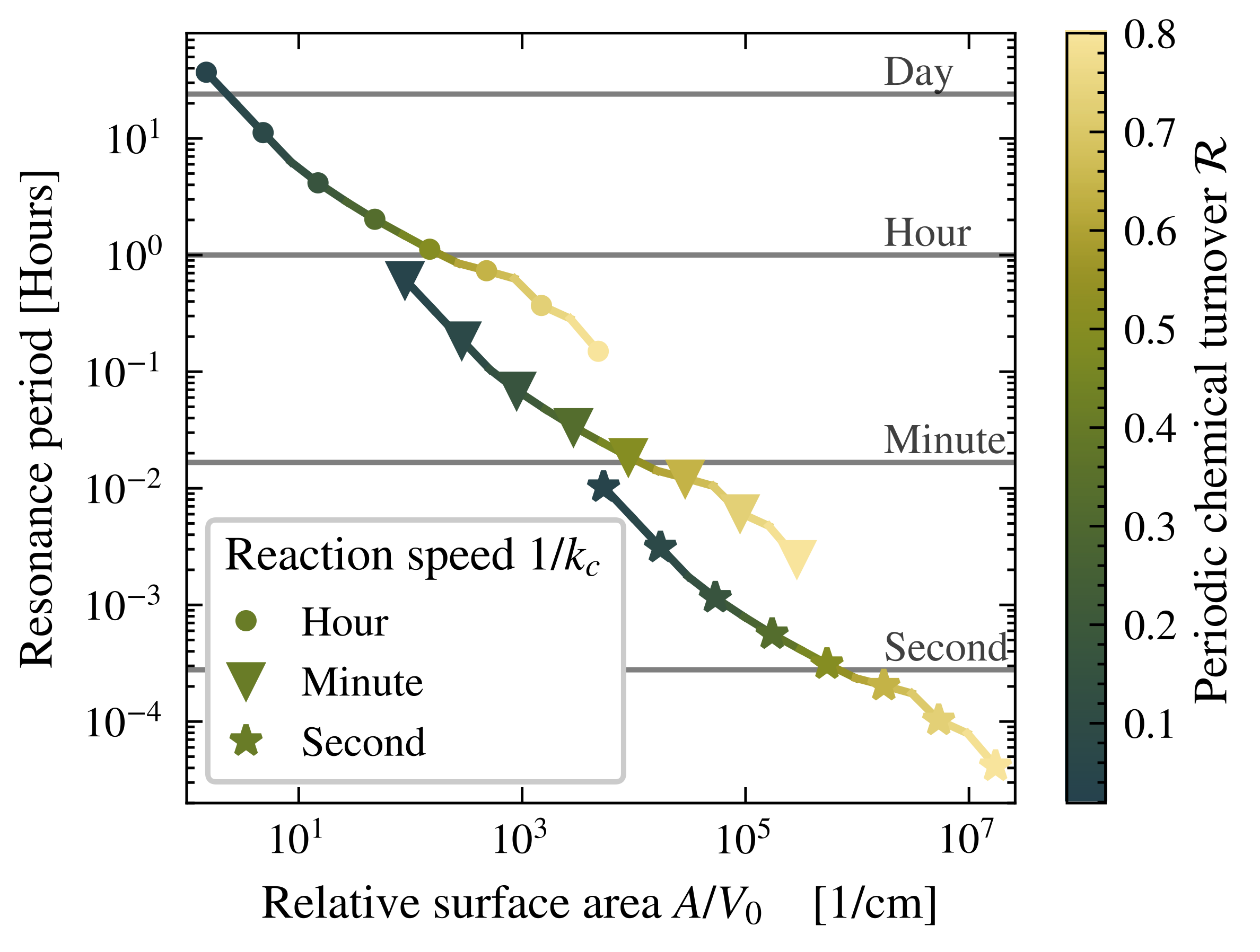}}}
     \caption{\textbf{Resonance behavior determined by system's geometry:}
     The resonance frequency of the periodic chemical turnover in Fig.~\ref{fig:5f} can be reached by varying the geometry of the mixture through its surface-to-volume ratio. Different reaction time-scales lead to different resonance frequencies, making the resonance behavior act as a selection mechanism. For this figure, the evaporation speed is {$5$~cm/day}.
     }\label{fig:6}
    \end{center}
\end{figure}

The resonance behavior provides a selection mechanism for specific chemical reactions. This selection is achieved by the speed-up of specific chemical reactions, while other reactions do not benefit from the wet-dry cycles. The selection conditions depend on the value of the evaporation flux coefficient $k_\text{e}$, which, however, 
does not change much for water for typical temperature and pressure oscillation on Earth~\cite{hisatake_evaporation_1993,heinen_evaporation_2019,xiong_evaporation_1991}.
A specific resonance frequency, and thus the selection of specific chemical reactions, can however be realized by varying the system's geometry, i.e., surface-to-volume ratio $A/V_0$ of the system (Fig.~\ref{fig:6}). 
Increasing the surface area to volume ratio from a puddle to a thin film increases the resonance frequency, selecting for faster chemical processes. 

\section*{Conclusion}
In this work, we develop a thermodynamic description for evaporation and condensation of non-dilute reacting mixtures and study the effect on chemical reactions. We find that kinetics of evaporation and condensation can significantly speed up chemical reactions. 
This speed-up stems from evaporation and condensation driving the system away from equilibrium. 
For wet-dry cycles, the turnover from substrate to product can be sped up even further.
We found that the cycling frequency strongly affects the rate of chemical reactions per cycle.
A key finding of our work is the existence of a resonance cycling frequency where turnover is maximal. 
This resonance behavior can act as a selection mechanism to only speed up specific reactions.

The mechanism for how evaporation and condensation for constant and cycling reservoirs affect chemical reactions relies on effectively lowering the free energy barrier $\Delta E= \mu_{A^*}^0 - \mu_A^0 + k_BT\log{(\gamma_{A^*}/\gamma_A)}$ for a reaction to occur (Fig.~\ref{fig:appendix1}(a), and Eq. \eqref{eq:1}, \eqref{eq:2}).
By reducing the amount of solvent through evaporation, a solvophilic component will interact less with the solvent and more with the other components, making it less energetically favorable to be solvophilic. 
Likewise, solvophobic substrates interacting more with the solvent makes it less energetically favorable to be solvophobic.
In both cases, the substrate chemical potential $\mu_A$ increases, thereby reducing the free energy barrier $\Delta E$.
For realistic thermodynamic and kinetic parameters of chemical compounds in aqueous solutions~\cite{clary_fast_1990,kok_prediction_1982,russell_characterization_2015,zwicker_growth_2017,hisatake_evaporation_1993,xiong_evaporation_1991}, speed-ups of chemical reactions by more than a factor 10 are possible through wet-dry cycles. Importantly, if assuming dilute chemistry $(\chi_{ij}=0)$, no speed-up or resonance would appear; the effect of wet-dry cycles results from dense chemistry interactions.
This pronounced effect on chemical reactions suggests a strong contender as a naturally occurring process analogous to enzymes. 
However, while enzymes speed up reactions by lowering
the kinetic contributions of the energetic barrier between the substrate and the product,
evaporation/condensation can increase the chemical potential of the substrate $\mu_A$ and lower the chemical potential of the product.
As a result, the speed of reactions can get enhanced (Fig.~\ref{fig:1a}).

Most chemical reactions in biology or chemical engineering involve enzymes or catalysts to enhance the speed of reactions~\cite{singh_catalysis_2014,werner_ionic_2010,blaser_role_1999,alberty:2003,srere_complexes_1987,underkofler_production_1958}. 
Subjecting such catalyzed reactions to wet-dry cycles could speed up the reactions even further. In particular, 
active sites of enzymes are typically solvent-depleted~\cite{zaks_effect_1988,brogan_enzyme_2014,helms_hydration_1998,klibanov_why_1997} implying that 
enzymatic activity is enhanced in 
the dry state of the cycle. 
Pronounced enhancement is expected when changing the time-dependent reservoir chemical potential from symmetric wet-dry cycles (Eq.~\eqref{eq:oscillations}) to strongly asymmetric cycles with long dry and short wet periods. 

In our work, the change in chemical reactions stems from the non-dilute conditions and the interactions among reacting components.  
Interestingly, 
phase separation did not qualitatively alter the speed-up or resonance phenomena observed. 
This might arise from our choice of equal kinetic coefficients in each phase. 
By including phase-dependent reaction coefficients or explicitly accounting for reactions on the interface between the two phases, we expect that phase separation can alter the resonance behavior.

In our work, we have focused on how chemical reactions are affected by wet-dry cycles in ternary mixtures. 
However, the theoretical framework provided can be applied to an arbitrary number of components with higher-order chemical reactions or reaction networks, as outlined in the theory section.
For such complex reaction schemes, we expect that the speed up of chemical reactions gets even more pronounced as the reaction rates become proportional to a product of all the substrate volume fractions (Eq. \eqref{eq:r_alpha}). This allows wet-dry cycles to affect chemical reaction rates even when assuming dilute chemistry \cite{higgs_effect_2016}. Moreover, the resonance behavior should persist as it arises from a generic coupling between reservoir driving and reaction rates.    
Exploiting this resonance behavior in reacting mixtures with many components  poses challenges at the interface between theory and experiments, which include the proper characterization of interaction parameters and kinetic rates.

Our findings suggest that wet-dry cycles could have played a vital role in speeding up prebiotic chemistry on the early Earth, where biological enzymes were absent.
During early Earth settings, systems were likely exposed to repeated wet-dry cycles. The reservoir oscillations can originate from weather changes such as temperature, humidity, or pressure~\citep{mulkidjanian_origin_2012,damer_coupled_2015}, or other phenomena such as salt deliquescence~\citep{campbell_prebiotic_2019}, or freeze-thaw cycles~\citep{vlassov_ligation_2004, le_vay_enhanced_2021,mutschler_freezethaw_2015}. Alternatively, oscillations in the solvent can be found in foams~\cite{tekin_prebiotic_2022}, or porous rock containing trapped gas-bubbles~\cite{ianeselli_non-equilibrium_2022,matreux_heat_2021}. 

An important implication of our work is that the resonance behavior in the frequency of wet-dry cycles provides a selection mechanism for chemical reactions.
Since the system's geometry determines the resonance frequency,
different chemical reactions are selected in a puddle versus a mesoscopic aqueous droplet. 
This suggests that pores of different sizes subject to wet-dry cycles can provide a setting for chemical selection and evolution at the molecular origin of life. 

\section*{Acknowledgement}
We thank Evan Spruijt and Iris Smokers for discussions on the experimentally observed effects of wet-dry cycles on chemical reactions and acknowledge their feedback on the manuscript. We also thank Sudarshana Laha for stimulating discussions and Hidde Vuijk, Samuel Gomez, and Gaetano Granatelli for feedback on the manuscript. Figure~\ref{fig:1} b,c was created using BioRender \cite{biorender}.
C.\ Weber acknowledges the European Research Council (ERC) under the European Union's Horizon 2020 research and innovation program (Fuelled Life,  Grant Number 949021) for financial support.

\appendix

\section{Validity of phase equilibrium  and parameter choices}\label{seq:limits}

Our work is valid for systems for which phase equilibrium approximately holds at each time during the kinetics.
Note that at phase equilibrium, the volume fractions in each phase are spatially homogeneous. 
For phase equilibrium to hold in the presence of other kinetic processes such as evaporation/condensation with a rate coefficients $k_e$ and chemical reactions with a rate coefficient $k_c$, such processes have to be slow compared to diffusion:
\begin{equation}
    k_e,\,\,k_c,\,\,\Omega \ll \frac{D_{\text{min}}}{l^2} \, , \label{eq:asump}
\end{equation}
where $l$ is the characteristic system length, and $D_{\text{min}}$ is the smallest diffusion coefficient out of all the components. Assuming phase equilibrium at each time of the kinetics also implies that the  nucleation of coexisting phases is not a rate limiting step and the effects of having a single droplet or an emulsion of the same total condensed volume are negligible. And indeed, nucleation for many phase separating polymeric systems shows for fast (often seconds or less) formation of mesoscopically sized condensed phases~\cite{bergmann_evolution_2022} and the effect of droplet number on composition is typically only relevant for droplets around the critical nucleation radius~\cite{weber_physics_2019}.

To see whether the condition in Eq.~\eqref{eq:asump} can be fulfilled for chemical systems we consider typical physico-chemical values for diffusion coefficients and chemical reaction rates. In water, small reacting molecules  often have a diffusion coefficient around $100\,\mu\text{m}^2/\text{s}$. When considering chemical reaction occurring with rates around $\text{min}^{-1}$,  the condition for phase equilibrium~\eqref{eq:asump} is satisfied for system sizes of $0.1$ mm or smaller. 
For larger system sizes,  gradients in the system should be taken into account by using a sharp interface model to calculate diffusive fluxes driven spatial inhomogeneities~\citep{bauermann_energy_2022}.

We have considered parameters consistent with the validity of the phase equilibrium condition~\eqref{eq:asump} during the wet-dry cycles and the reaction kinetics. The parameters used are listed in Table~\ref{tab:params}. In Fig. \ref{fig:3}, we span different interaction parameters to understand what class of interactions results in a kinetic speed-up. Our finding that turnover of solvophilic substrates is enhanced during drying agrees with the interactions of substrates used in phosphorylation reactions~\cite{maguire_physicochemical_2021,morasch_heated_2019}, where speed-up through wet-dry cycles also have been observed. The parameters in interactions used for Fig. \ref{fig:2}, \ref{fig:3}, \ref{fig:4} are motivated by systems with similar phase behaviour~\cite{fares_impact_2020,fritsch_local_2021,maguire_physicochemical_2021,bergmann_liquid_2023}. Moreover, also purified proteins show similar interactions~\cite{alberti_users_2018,arnold_phase_2007}.

We note that evaporating a major fraction of the solvent can strongly decrease the diffusion coefficient $D_\text{min}$, potentially violating the
condition~\eqref{eq:asump} as reacting compounds can become more crowded. However, 
removing the solvent also decreases the system's size $l$, potentially maintaining the validity of the condition above throughout the evaporation process.  


\begin{table*}[]
 \caption{The interaction strengths $\chi_{ij}$, internal energies $\omega_i$, and reservoir chemical potential values $\mu_C^r$ used for all figures is specified here. The $\Delta$ in the reservoir chemical potential means change relative to the initial chemical potential. For all figures in this work, we have chosen $\nu_i=1$, $k_c=1$, $V(t=0)=1$, and $\omega_C=0$. \label{tab:params}}
\begin{tabular}{@{}lllllllll@{}}
\toprule
Fig. $\qquad$& $\chi_{AC}/k_BT\quad$ & $\chi_{BC}/k_BT\quad$ & $\chi_{AB}/k_BT\quad$ & $\omega_A/k_BT\quad$ & $\omega_B/k_BT\quad$  & $\mu_C^r/k_BT$ \\ \midrule
\ref{fig:2a}     &    3.0      &    0.6      &    -0.6            &  -0.2      &   0.1   & (-0.2, -0.5, -0.9) \\
\ref{fig:2b} i   &    3.0      &    0.6      &   -0.6      &  -0.2        &   0.1   & -0.2 \\
\ref{fig:2b} ii  &    2.84     &    2.04     &   -0.6      &  -0.2      &   0.1      & -0.15 \\
\ref{fig:2b} iii &    1.72     &    1.72     &   -2.19     &  -0.1      &  0.4      & -0.2 \\
\ref{fig:2b} iv  &   -0.60     &    0.60     &    3.0      &  -0.2        &   0.1   & -1.40        \\
\ref{fig:3} a,b      &  3.0      &    0.6      &   -0.6      &  -0.2       &   0.1      & -0.5 \\
\ref{fig:4} a,b  &  3           &    -3      &      2       &   -10       &   0.0      &   -10        \\
\ref{fig:4}c  &  --          &    --      &     --       &   0.0     &   -6.0            &   $\Delta 2$       \\
\ref{fig:4} d,e  &  2          &    3.5      &      -2       &   -10     &   0.0            &  $ 2.77$       \\
\ref{fig:4}f  &  --          &    --      &     --       &  -6.0     &   0.0            &   $-\Delta 5$       \\
\ref{fig:5} a,b,c &  3         &    -0.6      &     0.6        &   0       &    -6.0            &   $\langle-0.3  \rangle$ \\
\ref{fig:5} d,e,f &  3         &    -0.6      &     0.6        &   0.4     &   1.8            &   $\langle-0.538\rangle$      \\
\bottomrule
\end{tabular}
\end{table*}

\section{Unidirectional chemical reactions and analytic scaling relation}\label{sec:unidir}
Unidirectional chemical reactions are achieved by choosing a large difference in the internal energies between the product and substrate
\begin{equation}
    \omega_\text{product} \ll \omega_\text{substrate} \, . \label{eq:uni_condition}
\end{equation}
In turn, the chemical reaction becomes dominated by the direction from substrate to product, where the backward pathway is exponentially damped by the factor $(\omega_\text{product} - \omega_\text{substrate})/k_\text{B}T $. The chemical reaction in equation \eqref{eq:chem_react} can, under these conditions, be approximately written as
\begin{equation}
     r_{A\rightharpoonup B} - r_{A \leftharpoondown B} \approx \tilde{k}_{c}\phi_A\exp{\frac{\tilde{\mu}_A}{k_\text{B}T}} \, ,
\end{equation}
when $A$ is acting as the substrate and $B$ as the product. We have defined $\tilde{k_c}\equiv k_c\exp{\omega_A/k_\text{B}T}$, and $\tilde{\mu}_A$ represents the non-entropic contribution to the chemical potential of component $A$, defined as
\begin{equation}
    \frac{\tilde{\mu}_A}{\nu_A} = \phi_B \chi_{AB} + \phi_C\chi_{AC} + p - \Lambda + k_\text{B}T/\nu_A\, . \label{eq:chi_speedup}
\end{equation}
To speed up the chemical reaction, this quantity should be maximized. During evaporation, we reduce $\phi_C$, increasing $\phi_A$ and $\phi_B$. A negative value of $\chi_{AC}$ increases the substrate's internal energy. While for condensation, $\phi_C$ is increased relative to $\phi_A$, such that the chemical potential is increased for positive values of $\chi_{AC}$. Though the interaction between $\chi_{AB}$ might appear as important as $\chi_{AC}$, the volume fraction of the product will generally vary less during evaporation/condensation. However, it will become important towards the later part of the turnover process. Importantly, $\Lambda$, defined in Eq. \eqref{eq:lambda}, also depends on the volume fractions and will vary during evaporation/condensation. The changes in $\Lambda$ through evaporation/condensation will still contribute less to the overall speed-up as its interaction terms contribute only as second order in the $\phi$'s.\par 
The reaction coefficients $k_{c,\alpha}^{\RN{1}/\RN{2}}$ can generally depend on the volume fractions and differ in each phase. However, for simplicity, we treat it as constant and equal in the two phases.

\section{Phase equilibrium constraint during reaction and wet-dry cycle kinetics}\label{sec:PS_kin}

In our work, we consider the kinetics of chemical reactions and wet-dry cycles, where we focus on spatially homogeneous chemical potential. Either the system is spatially homogeneous in terms of composition, or phase-separated  with phase equilibrium between phase I and II. 
Below we review how the constraint of phase equilibrium is accounted for in the numerical solutions; more details can be found in Ref.~\cite{bauermann_chemical_2022}.

To evolve the phase equilibrium constraint, $\mu_i^\RN{1}=\mu_i^\RN{2}$, and thereby calculate the volume fractions in the two phases   (Eq.~\eqref{eq:kinetics}), we use
\begin{equation} 
    \partial_t\mu_i^\RN{1} = \partial_t\mu_i^\RN{2} \, . \label{eq:phase_coex}
\end{equation}
Using the product rule, this can be rewritten to
\begin{equation}
    \partial_t\mu_i^{\RN{1}/\RN{2}} = \sum_{j=0}^M \pdv{\mu_i^{\RN{1}/\RN{2}}}{\phi_j^{\RN{1}/\RN{2}}}\pdv{\phi_j^{\RN{1}/\RN{2}}}{t}.
\end{equation}
Inserting the time-derivative of $\phi_i^{\RN{1}/\RN{2}}$ from Eq.~\eqref{eq:kinetics}, Eq.~\eqref{eq:phase_coex} becomes
\begin{align}
    &\sum_{j=0}^M \pdv{\mu_i^\RN{1}}{\phi_j^\RN{1}}\left(r_j^{\RN{1}} - j_j^{\RN{1}} + \frac{\nu_j}{V^{\RN{1}}}h_j^{\RN{1}}- \phi_j^{\RN{1}} \sum_{k=0}^M \left(\frac{\nu_kh_k^{\RN{1}}}{V^\RN{1}} - j_k^{\RN{1}}\right)\right) \label{eq:find_ji} \\
    = &\sum_{j=0}^M \pdv{\mu_i^\RN{2}}{\phi_j^\RN{2}}\left(r_j^{\RN{2}} - j_j^{\RN{2}} + \frac{\nu_j}{V^{\RN{2}}}h_j^{\RN{2}}- \phi_j^{\RN{2}} \sum_{k=0}^M \left(\frac{\nu_kh_k^{\RN{2}}}{V^\RN{2}} - j_k^{\RN{2}}\right)\right) \, ,\nonumber
\end{align}
where $r_j^{\RN{1}/\RN{2}}$, $j_j^{\RN{1}/\RN{2}}$, and $h_j^{\RN{1}/\RN{2}}$ are the reaction, diffusive and evaporation/condensation fluxes.  
The conservation of particle number through the interphase implies
\begin{equation}
    V^\RN{1}j_i^\RN{1} = - V^\RN{2}j_i^\RN{2},
\end{equation}
and can be plugged into Eq.~\eqref{eq:find_ji}. Thus, our $(M+1)$ unknown $j_i^\RN{1}$ can be calculated by solving the $(M+1)$ coupled algebraic equation in Eq.~\eqref{eq:find_ji}. 
\vspace{1cm}

\section{Role of transition states}\label{sec:transition_states}

\begin{figure}[b]
    \centering
     \makebox[\textwidth][c]{
    \includegraphics[width=\textwidth]{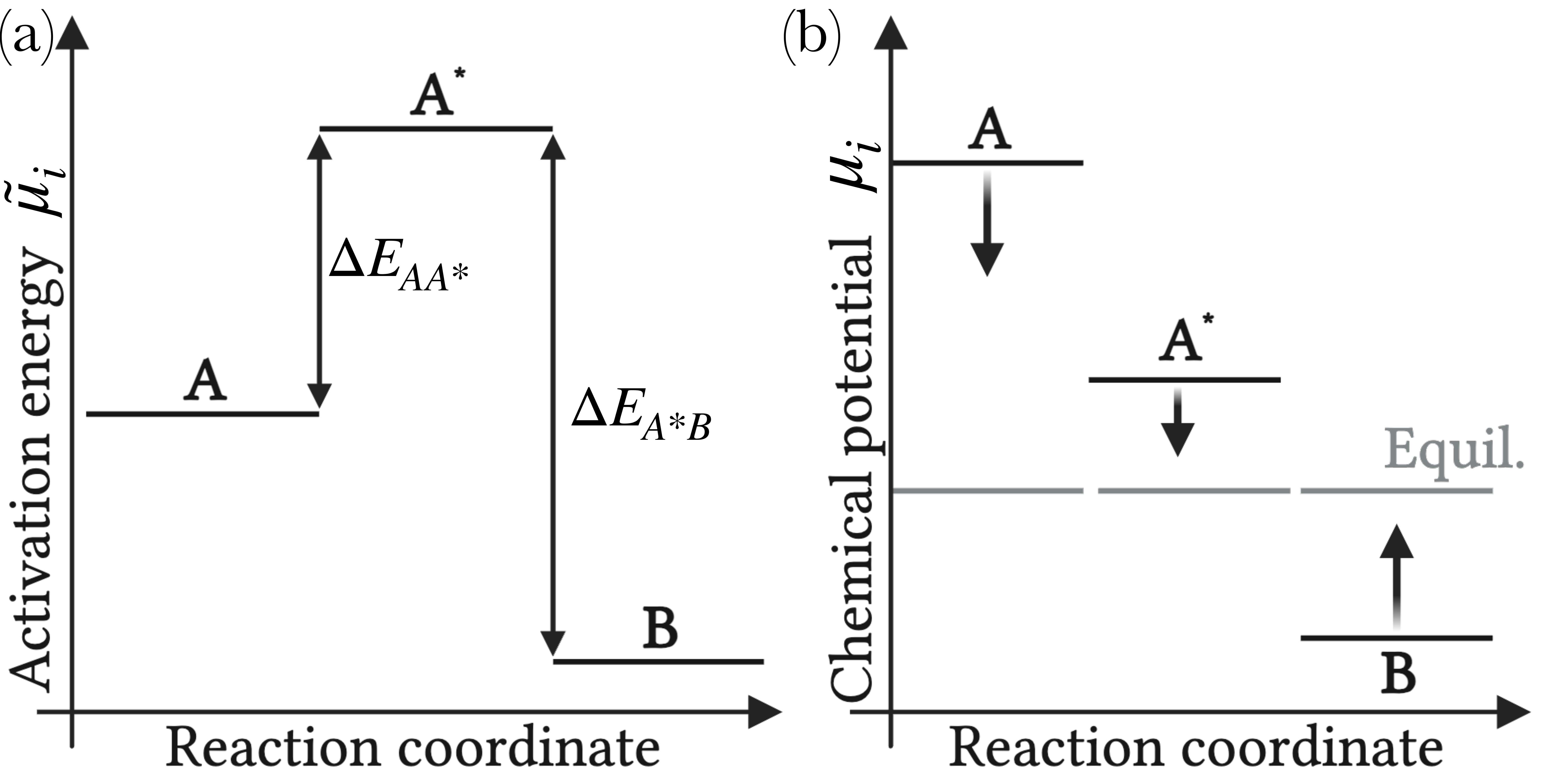}}
    \caption{\textbf{Chemical reactions with a transition state:}
    The transition state $A^*$ between the reacting components $A$ and $B$ gives rise to an energetic barrier for the reaction $A\rightleftharpoons B$. 
    (a) The energy landscape is set by  the chemical potential $\tilde{\mu}_i\equiv \mu_i^0 + k_BT\log{\gamma_i}$, where $\mu_i^0$ are the reference chemical potentials and $\gamma_i$ is the activity coefficient that includes the interactions among the non-dilute components. The activation energy  for the reaction $A\rightleftharpoons A^*$ to occur is $\Delta E_{AA^*} = \mu_{A^*}^0 -\mu_A^0 + k_BT\log{\gamma_{A^*}/\gamma_{A}}$, and $\Delta E_{A^*B} = \mu_{A^*}^0 -\mu_B^0 + k_BT\log{(\gamma_{A^*}/\gamma_{B})}$ for the reaction $A^*\rightleftharpoons B$. For large energetic barriers, the transition state will be both dilute and short-lived, yielding a separation of time scales. 
    (b) The chemical potentials $\mu_i$ change in time and become equal at chemical equilibrium. Moreover, for a quasi-static transition state ($\partial_t \phi_{A^*} \simeq 0$), the chemical potential of $A^*$ corresponds to a weighted average
    between $\mu_A$ and $\mu_B$ at each time of the kinetics (Eq.~\eqref{eq:slaveA*}). 
    }
\label{fig:appendix1}
\end{figure}

In this section, we discuss the role of  transition states in the kinetics of chemical reactions  when subject to wet-dry cycles.
To this end, we introduce the transition state $A^*$, which is  associated with  energetic barriers, typically referred to as  activation energies, relative to the states $A$ and $B$, $\Delta E_{AA^*}$ and $\Delta E_{A^*B}$, respectively  (Fig.~\ref{fig:appendix1}(a)).
For the sake of illustration, we restrict ourselves to homogeneous mixtures but allow the system to be non-dilute. Due to the change in composition during wet-dry cycles, the activation energies, in general, vary during the chemical  kinetics. 
Here, we show that the effects of transition state $A^*$ on the kinetics are negligible if the activation energies heights are in the order of a few $k_BT$ or larger.
Such large energetic barriers will lead to a dilution of the transition state corresponding to low volume fractions $\phi_{A^*}$.

The chemical reaction rates for a chemical system with the substrate $A$, transition state $A^*$ (Fig.~\ref{fig:appendix1}), and a product $B$ are given as:

\begin{align}
    r_{A\rightleftharpoons A^*} &= k_{AA^*} \left[\exp{\frac{\mu_A}{k_BT}} - \exp{\frac{\mu_{A^*}}{k_BT}} \right] \, , \label{eq:r1}\\
    r_{A^* \rightleftharpoons B} &= k_{A^*B} \left[\exp{\frac{\mu_{A^*}}{k_BT}} - \exp{\frac{\mu_{B}}{k_BT}} \right]\, . \label{eq:r2}
\end{align}
In a homogeneous system, they govern the rate of change  of the volume fraction of the three  components:
\begin{align}
\begin{split}
    \partial_t\phi_A     &= -r_{A\rightleftharpoons A^*}           \, ,\\
    \partial_t\phi_{A^*} &= r_{A\rightleftharpoons A^*} - r_{A^* \rightleftharpoons B}\, , \label{eq:pdv_phia*}\\
    \partial_t\phi_B     &= r_{A^* \rightleftharpoons B}  \, .
\end{split}
\end{align}
Solving these equations, we find that for large enough barriers,
$\Delta E_{AA^*}$ and $\Delta E_{A^*B}$, the transition state quickly relaxes to quasi-static conditions ($\partial_t\phi_{A^*} \simeq 0$). 
To understand this analytically, we consider an initial state $\phi_A(0)=\phi_A^0$, $\phi_{A^*}(0)=0$, and  $\phi_{B}(0)=0$, and consider that $\phi_A(t)\simeq \phi_A(0)$ is at excess at early times. 
Thus, one can write the time-derivative of $A^*$ (\ref{eq:r1}-\ref{eq:pdv_phia*}) as
\begin{align}
\partial_t\phi_{A^*} \simeq -&\left(k_{AA^*}+k_{A^*B}\right)\phi_{A^*}\exp{\frac{\mu_{A^*}^0}{k_BT} + \log{\gamma_{A^*}}} \nonumber \\ + &k_{AA^*}\phi_A^0\exp{\frac{\mu_A^0}{k_BT} + \log{\gamma_A}}. \label{eq:phiastar_equation}
\end{align}
To achieve these expressions, we have rewritten the chemical potentials, $\mu_{i} = \mu_i^0(T, p) + k_BT\log{\left(\phi_i\gamma_i\right)}$, where 
\begin{align}
    \log{\gamma_i} &=  \nu_i \sum_{j=0}^M \left( \chi_{ij}\phi_j - k_BT\frac{\phi_j}{\nu_j} - \sum_{k=0}^M\frac{\chi_{jk}\phi_j\phi_k}{2}\right) \, , \label{eq:1}\\
     \mu_i^0 &= k_BT + \omega_i + \nu_i p \, \label{eq:2}.
\end{align}
Here, $\gamma_i$ are the activity coefficients, including the effects of interactions, and $ \mu_i^0$ are the reference chemical potentials.
The chemical potential $(\mu_i^0+\log{\gamma_i})$ describes  the energy landscape that a molecule experiences while reacting between $A$ and $B$. It allows us to define activation energies for non-dilute systems, 
$\Delta E_{AA^*} = \mu_{A^*}^0 -\mu_A^0 + k_BT\log{\gamma_{A^*}/\gamma_{A}}$ and $\Delta E_{A^*B} = \mu_{A^*}^0 -\mu_B^0 + k_BT\log{(\gamma_{A^*}/\gamma_{B})}$. 
(Fig.~\ref{fig:appendix1}(a)).
Note that such activation energies, in general, depend on composition and are only constant for dilute mixtures.

The solution to Eq. \eqref{eq:phiastar_equation} at early times reads: 
\begin{equation}
    \phi_{A^*}(t) = \Phi_{A^*}\left(1-\exp{-t/\tau}\right)\, , \label{eq:solution_A*}
\end{equation}
with the characteristic relaxation time 
\begin{equation}
    \tau = \frac{1}{k_{AA^*} + k_{A^*B}}\exp{-\frac{\mu_{A^*}^0}{k_BT} - \log{\gamma_{A^*}}}\, , \label{eq:tau}
\end{equation}
and the early-time plateau value of the transition state 
\begin{align}
    \Phi_{A^*} &\simeq \frac{k_{AA^*}}{k_{AA^*}+k_{A^*B}}\phi_A^0\exp{-\frac{\Delta E_{AA^*}}{k_BT}}\, , \label{eq:plateau}
\end{align}
where $\Delta E_{AA^*}$ is the activation energy  between $A$ and $A^*$ as depicted in Fig.~\ref{fig:appendix1}.

From the expression, we observe that the occupation of the transition state decreases exponentially with a characteristic time-scale $\tau$.
This time scale has an
exponential dependence on the activation energy  $\Delta E_{AA^*}$. 
Thus, for larger barriers, the relaxation of the transition state occurs on a time scale much faster $k_{AA^*}$ and $k_{A^*B}$. After this relaxation, the kinetics of the transition state is quasi-statically slaved to the slow changes of $A$ and $B$. Thus, the transition state satisfies   $\partial_t \phi_{A^*}\simeq 0$. On such time scales, i.e,  $t\simeq k_{AA^*}^{-1},k_{A^*B}^{-1}$,  $\Phi_{A^*}$ and $\phi_A$ change concomitantly in time leading to the population of the $B$ state.

\begin{figure}[b]
    \centering
     \makebox[\textwidth][c]{
    \includegraphics[width=0.9\textwidth]{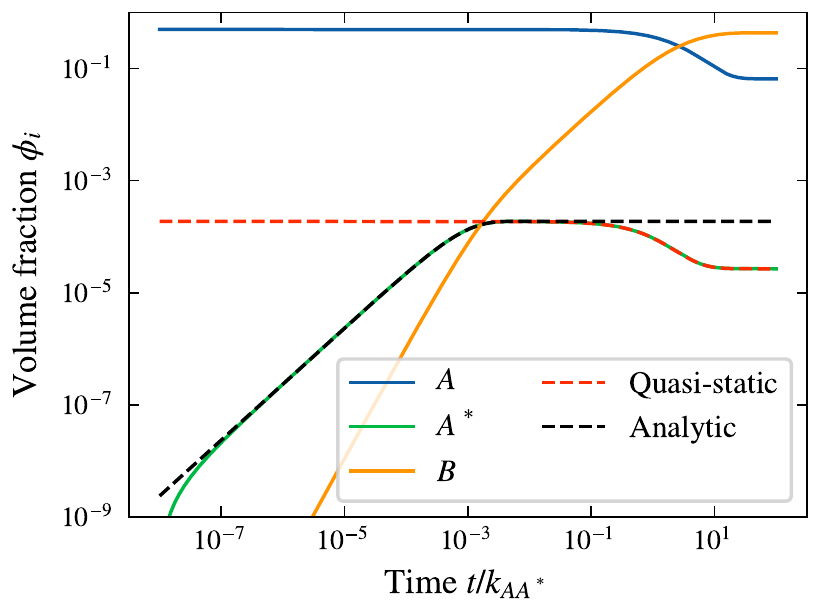}}
    \caption{\textbf{Analytic equilibration of the transition state is fast:} The full numerical solution is seen to agree with the approximate analytic solution in equation \eqref{eq:solution_A*}. The transition state's relaxation time scale is faster than for $A$ and $B$, and remains dilute throughout.
    }
   \label{fig:appendix2}
\end{figure}

We note that large activation energies $\Delta E_{AA^*}$ and $\Delta E_{A^*B}$ also imply that the transition state $A^*$ gets diluted, which is reflected in the exponential dependence of $\Phi_{A^*}$ on the barrier height (Eq.~\eqref{eq:plateau}). 
In other words, transition states with large energetic barriers lead to a separation of time scales and a dilution of the transition evolving accordingly to a quasi-static kinetics~\cite{van_kampen_elimination_1985,janssen_elimination_1989}.
This statement is confirmed by the excellent agreement between the approximate analytic expression \eqref{eq:solution_A*} and the full numerical solution of equation \eqref{eq:pdv_phia*} (Fig.~\ref{fig:appendix2}). 
At later time scales when $A$ is converted to $B$, the plateau value of $\Phi_{A^*}$ is slowly changing and thus deviates from the analytic value Eq.~\eqref{eq:plateau}.
However, solving Eq.~\eqref{eq:pdv_phia*} using a quasi-static approximation ($\partial_t \phi_{A^*} \simeq 0$)
agree extremely well with the full numerical solution of Eq. \eqref{eq:pdv_phia*} at later times.
\par 

We conclude that for large barriers and large times $t>\tau$, the quasi-static condition is well satisfied.
Using Equations (\ref{eq:r1}-\ref{eq:pdv_phia*}), we find that the chemical potential of the transition state corresponds to a weighted average of the chemical potentials of $A$ and $B$:
\begin{equation}\label{eq:slaveA*}
    \exp{\frac{\mu_{A^*}}{k_BT}} = \frac{k_{AA^*}\exp{\frac{\mu_A}{k_BT}} + k_{A^*B}\exp{\frac{\mu_B}{k_BT}}}{k_{AA^*} + k_{A^*B}} \, . 
\end{equation}
This equation can be solved for $\phi_{A^*}$ at any time point of the later-time kinetics.

By substituting Eq.~\eqref{eq:slaveA*} into Eq.~\eqref{eq:r1}, we find
\begin{equation}
    r_{A\rightleftharpoons A^*} = \frac{k_{AA^*}k_{A^*B}}{k_{AA^*} + k_{A^*B}}\left[\exp{\frac{\mu_A}{k_BT}} - \exp{\frac{\mu_{B}}{k_BT}} \right] \, ,
\end{equation}
and  $ r_{A\rightleftharpoons A^*} = r_{A^* \rightleftharpoons B}$. Comparing to Eq.~\eqref{eq:chem_react} and Eq.~\eqref{eq:r_alpha}, we see that the expression is equivalent to a system with a single mono-molecular chemical reaction $\sigma_{i\alpha}^\rightharpoonup = \delta_{iA}\delta_{\alpha1}$, $\sigma_{i\alpha}^\leftharpoondown = \delta_{iB}\delta_{\alpha1}$, and $k_{c,\alpha}=k_c\delta_{\alpha,1}$. Thus, we introduce an effective modified chemical reaction rate coefficient
\begin{equation}
    k_c \equiv \frac{k_{AA^*}k_{A^*B}}{k_{AA^*} + k_{A^*B}} \, . \label{eq:effectiv_rate}
\end{equation}
We conclude that for large barrier heights, the kinetics of $A$ and $B$ separated by a transition state $A^*$ can be mapped on a reduced chemical system without a transition state with an effective reaction rate $k_c$ (Eq \eqref{eq:effectiv_rate}).\par 
\begin{figure}[b]
    \centering
     \makebox[\textwidth][c]{
    \includegraphics[width=0.9\textwidth]{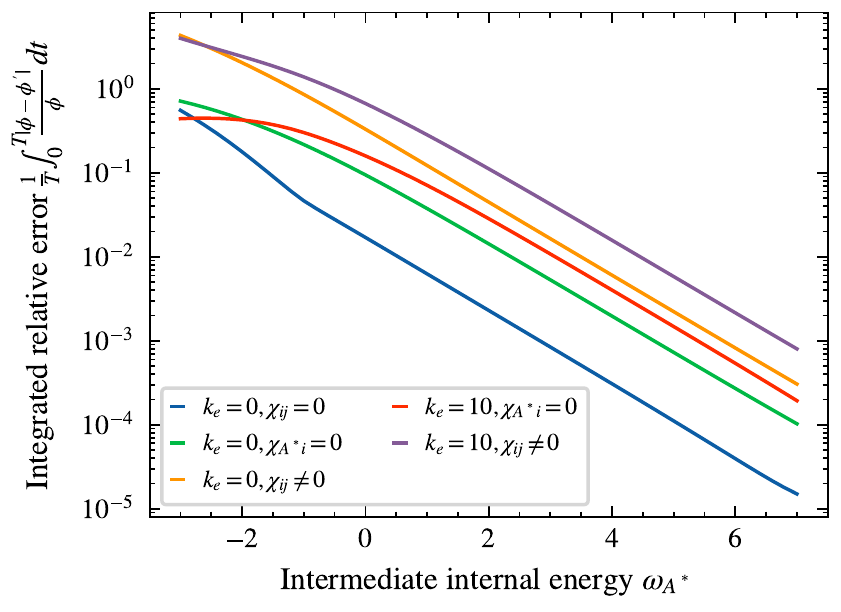}}
    \caption{\textbf{Relative error with and without a transition state:} The integrated relative difference between the substrate volume fraction with and without a transition state decreases exponentially with the barrier height. By including interactions and/or evaporation/condensation, the required barrier height to achieve a given error increases. For this graph $\omega_A=0$ and $\omega_B=-3k_BT$.
    }
   \label{fig:appendix3}
\end{figure}
This reduction is also valid when the system is subject to evaporation if the time scales of evaporation/condensation $k_e^{-1}$, reaction $k_c^{-1}$ and $\Omega^{-1}$ are slow compared to the time scale of the transition state relaxation $\tau$ \eqref{eq:tau}:
\begin{equation}
    \tau \ll {k_e}^{-1} , {k_c}^{-1}, \Omega^{-1} \, .
\end{equation}
This statement is supported by Fig.~\ref{fig:appendix3}, where the integrated relative difference between the substrate volume fraction for the kinetics with and without the transition state decreases exponentially with the barrier height.\par

The elimination of the kinetic equation for the transition state $A^*$ does not imply that the effective rate $k_c$ (Eq.~\eqref{eq:effectiv_rate}) was independent of the transition state.
Using Kramer's transition state theory~\cite{gardiner_handbook_1985}, the kinetic rate coefficient $k_{AA^*}$ and $k_{A^*B}$ exponentially depends on the respective activation energies:
\begin{align}\label{eq:Kramer_for_kinrates}
    k_{AA^*} &= w_{AA^*}\exp{-\frac{\Delta E_{AA^*}}{k_BT}} \, , \\
    k_{A^*B} &= w_{A^*B}\exp{-\frac{\Delta E_{A^*B}}{k_BT}}\, .
\end{align}
The above relationships not only encodes the effects of transition state $A^*$  for large activation energies, it also implies that such effects in general depend on the composition of all components due to the interaction among each other.  Yet, as long as the reference chemical potentials dominate the activation energies, i.e., they exceed the contribution related to the activity coefficients,
\begin{equation}
    \mu_{A^*}^0-\mu_{A}^0 \gg k_BT\log{\left( 
 \frac{\gamma_A}{\gamma_{A^*}} \right)} \, ,
\end{equation}
the kinetic rates  $ k_{AA^*} $ and $k_{A^*B}$ (Eq.~\eqref{eq:Kramer_for_kinrates}) are approximately composition-independent. In this case, the effective reaction coefficient (Eq.~\eqref{eq:effectiv_rate}) is approximately composition-independent, too.
\vspace{0.5cm}

\section{Gibbs' phase rule}\label{sec:Gibbs}
For a system of $\mathcal{C}$ non-solvent components, $\mathcal{N}$ external thermodynamic parameters, and $\mathcal{P}$ number of phases, the number of degrees of freedom $\mathcal{F}$ can be calculated from 
\begin{equation}
    \mathcal{F} = \mathcal{C} + \mathcal{N} - \mathcal{P}.
\end{equation}
The number of degrees of freedom gives the number of independent intensive variables uniquely defining the thermodynamic equilibrium. The number of components $\mathcal{C}$ is calculated as the number of components needed to specify the composition of all phases (incompressibility reduces $\mathcal{C}$ by one) minus the number of constraints between the concentrations (reduced by the number of chemical reactions). Furthermore, the evaporation/condensation equilibrium can be regarded as an external intensive variable defining the system. Such that the set $\{T, p, \mu_C^r\}$ or alternatively $\{T, \mu_C^r\}$, gives
\begin{equation}
    \mathcal{F}= \begin{cases}  5-\mathcal{P}  & \mbox{for} \,\quad \{T, p, \mu_C^r\} \\
                      4-\mathcal{P}  & \mbox{for} \,\quad \{T, \mu_C^r\} \end{cases}
\end{equation}
Therefore, two-phase coexistence in a system is defined as a volume in $\{T, p, \mu_C^r\}$-space and a plane in $\{T, \mu_C^r\}$-space, while the homogeneous system has an additional degree of freedom. For a fixed reservoir and a single chemical reaction, the degree of freedom becomes $\mathcal{F}=2-\mathcal{P}$, such that two-phase coexistence is only achievable for a single point in the phase diagram. This is consistent with the phase diagram in Fig.~\ref{fig:3c}.

\bibliography{citations_final.bib}

\section*{}
\newpage
\begin{figure}[b]
    \centering
     \makebox[\textwidth][c]{
    \includegraphics[width=8.25cm]{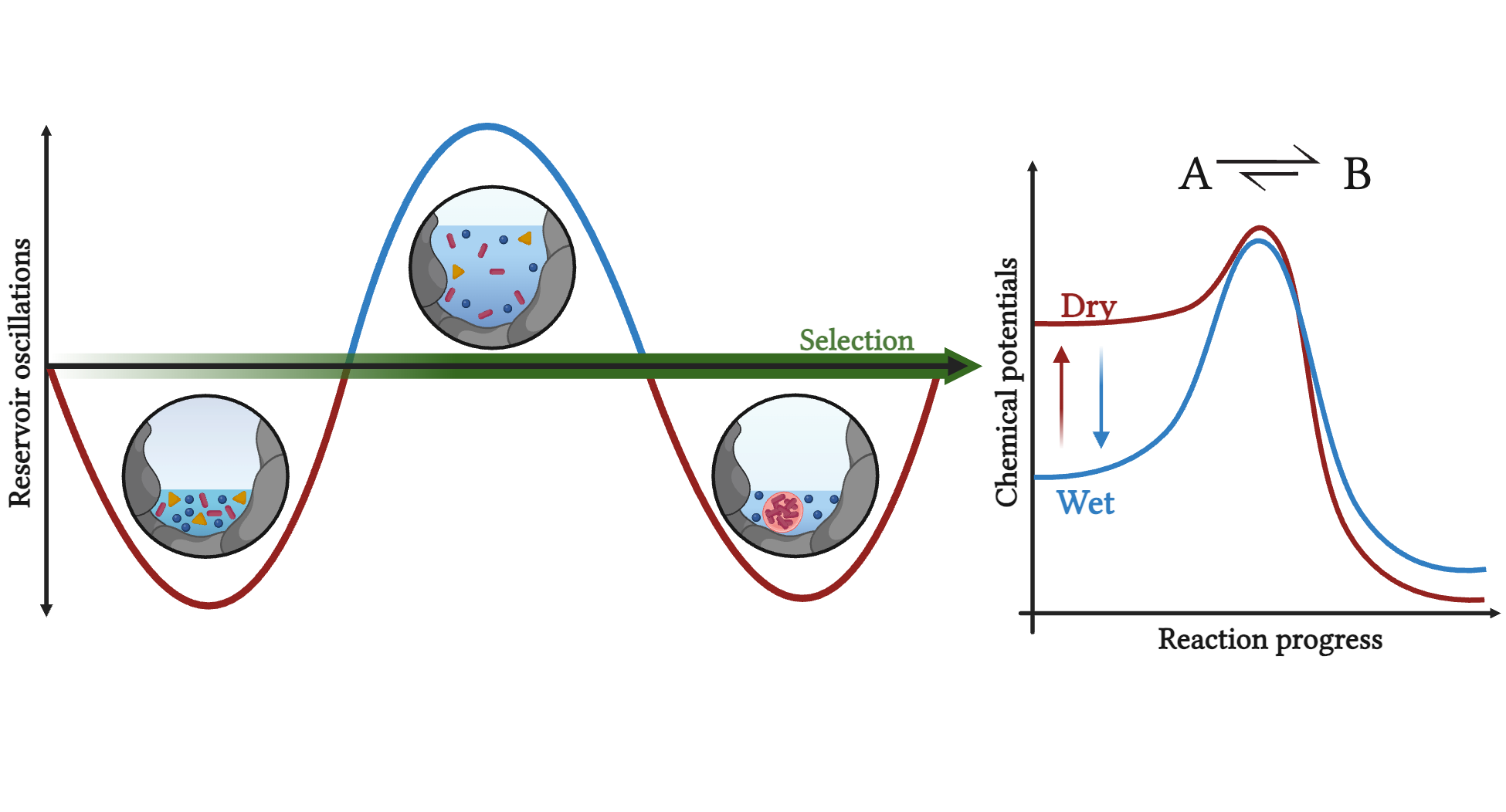}}
    \caption{TOC Graphic}
\end{figure}

\end{document}